\documentclass[twocolumn]{aastex7}

\newcommand{\kms}{km~s$^{-1}$\,}

\newcommand{\msun}{${\cal M}_\odot$\,}

\begin{document}

\renewcommand{\topfraction}{1.0}
\renewcommand{\bottomfraction}{1.0}
\renewcommand{\textfraction}{0.0}

\shorttitle{Speckle Interferometry at SOAR}

\title{Speckle Interferometry at SOAR in 2024 and 2025}

\author{Andrei Tokovinin}
\affil{Cerro Tololo Inter-American Observatory --- NFSs NOIRLab Casilla 603, La 
Serena, Chile}
\email{andrei.tokovinin@noirlab.edu}

\author{Brian D. Mason}
\affil{U.S.\ Naval Observatory, 3450 Massachusetts Ave., Washington, DC, USA}
\email{brian.d.mason.civ@us.navy.mil}

\author{Rene A. Mendez}
\affil{Universidad de Chile, Casilla 36-D, Santiago, Chile}
\email{rmendez@uchile.cl}
\author{Edgardo Costa}
\affil{Universidad de Chile, Casilla 36-D, Santiago, Chile}
\email{costa@das.uchile.cl}

\begin{abstract}
Results of speckle  interferometry observations at the  4.1 m SOuthern
Astrophysical Research (SOAR) telescope obtained during 2024--2025 are
presented.   We present 5316 measurements  of relative positions
and  magnitude differences  in 3532  pairs (including  524 unpublished
measures  made before  2024) with  median and  minimum separations  of
0\farcs19 and 12\,mas, respectively; non-resolutions of 1723 stars are
documented as  well. More than 400  pairs have been resolved  here for
the first  time and  not resolved  by Gaia; among  those are  222 TESS
objects  of interest,  46  inner subsystems  in  known wider  binaries
within 100\,pc, and 43 subdwarfs.  Positional measurements are used to
compute or  improve binary orbits;  elements of  202  orbits 
  with meaningful errors  are given  here,  while preliminary and
  tentative orbits are published elsewhere. Of special note are orbits
  with large and accurately measured eccentricties (e.g. $e=0.9866 \pm
  0.0014$  for   J13038$-$2035)  and   orbits  of   pre-main  sequence
  binaries.   Appendix contains  parameters of  86 binaries  used for
calibration of pixel scale and orientation.
\end{abstract} 

\section{Introduction}
\label{sec:intro}

 Speckle interferometry replaced visual micrometer observations of
  double stars in the last decades of the 20th century, improving both
  the accuracy and the angular resolution.  At that time, most speckle
  observations      were     made      by     the      CHARA     group
  \citep{McAlister1993,Mason2009,Mason2026} to monitor orbital motions
  and  to discover  new close  pairs. Nowadays,  solid-state detectors
  like  electron-multiplication (EM)  CCDs and  CMOS make  this method
  more practical and efficient.  Without being exhaustive, we can cite
  recent         works         by         \citet{Horch2017,Horch2021},
  \citet{Mitrofanova2021}, and \citet{Clark2024}.  Speckle instruments
  have  been used  extensively to  characterize duplicity  of exohosts
  \citep{Howell2021a,Lester2023}   and  to   survey  diverse   stellar
  populations    for   companions    \citep[e.g.    nearby    low-mass
    stars,][]{Janson2012,Vrijmoet2022}.  Speckle  interferometry is by
  far  the  largest source  of  high-resolution  astrometry of  binary
  stars,   compared   to    adaptive   optics   \citep{Mann2019}   and
  long-baseline interferometry \citep{Torres2022}.  Owing to its lower
  spatial resolution  and decade-long  time span,  the data  from Gaia
  \citep{Gaia1} cannot replace speckle interferometry.  

This paper  continues the series  of double-star measurements  made at
the 4.1 m SOuthern Astrophysical  Research (SOAR) telescope since 2008
with the  speckle camera,  HRCam.  Previous  results are  published by
\citet[][hereafter           TMH10]{TMH10}            and           in
\citep{SAM09,Hrt2012a,Tok2012a,TMH14,TMH15,SAM15,SAM17,
  SAM18,SAM19,SAM20,SAM21,SAM22,SAM23}.   Observations  reported  here
were   made  during   2024--2025;  we   also  give   some  unpublished
observations  made  before  2024.

The structure  and content of this  paper are similar to  the previous
papers  of this  series.  Section~\ref{sec:obs}  reviews all  speckle
programs that contributed to this  paper, the observing procedure, and
the    data    reduction.     The    results    are    presented    in
Section~\ref{sec:res} in the form of electronic tables archived by the
journal.  We also discuss new resolutions and present orbits resulting
from   this    data   set.     A   short    summary   is    given   in
Section~\ref{sec:sum}. The Appendix contains  data on 86 binaries used
for calibration of pixel scale and orientation. 

\section{Observations}
\label{sec:obs}

\subsection{Observing Programs}

The   observations  reported   here  were   obtained  with   the  {\it
  high-resolution camera} (HRCam)  --- a fast imager  designed to work
at the  4.1 m  SOAR telescope  \citep{HRCAM}. It  was used  to execute
several observing programs, some with common targets.  About 10 nights
per  year  were  used,  divided  almost  evenly  between    regular
allocations and additional engineering time (usually morning
hours on bright-Moon nights). A total of 21 individual observing runs,
ranging from a few hours to two nights, were executed in 2024--2025.

The allocated time was distributed between the following projects:
\begin{itemize}
\item
Multiple stars and orbits (3 nights, PI A.T.)

\item
Orbits and masses (4 nights, PI R.M.)

\item
Orbits of nearby M  dwarfs (one night, PI E.~Vrijmoet).   Most observations up
to 2025.1 are published by \citet{Vrijmoet2026}; they are
provided here for completeness.   The  first results  of  this
long-term effort are published by \citet{Vrijmoet2022}.

\item
Survey of nearby M dwarfs (two nights, PI M.~Leblanc). These data (mostly
non-resolutions) are not published here.

\item
TESS  follow-up (one  night  of  SOAR partner's  time  in 2024).   All
observations  are  given  here  and    posted  at  the  Exofop
site,\footnote{\url{https://exofop.ipac.caltech.edu/tess/}}     see     also
\citet{TESS,TESS2}. Previously unpublished measures of resolved TESS
objects of interest (TOIs)  made after 2021.09 are included in our
data table, see Section~\ref{sec:TESS}. 

\end{itemize}

The engineering  time was  used to  complement the  scheduled programs
(for example, to compensate weather and poor-seeing losses) and to observe
orbital or neglected  pairs. Additional details on  the HRCam programs
and their motivation can be found in \citet{SAM23}.  The HRCam results
are eventually ingested into the Washington Double Star (WDS) database
\citep{WDS}.

The proprietary period of the SOAR data is 18 months. To prevent loss
of information, we include in  the data tables the HRCam observations
made in 2021--2022 for the  projects led by K.~Franson and B.~Bowler
(nearby  stars  with  astrometric  acceleration)  and  a  search  for
resolved   subsystems   in   wide    binaries   from   the   list   of
\citet{Andrews2017} led by J.~Chanam\'e; we expect that these projects
will publish the HRCam data and  their analysis in the future. We also
include  here  recent observations  of  hierarchical  systems used  in
\citep{TRI25,CHI25,TRI26}   and   observations   of   doubly-eclipsing
quadruples \citep{Majewski2025}.

Speckle observations with  HRCam is an iterative  cycle beginning with
observing program  preparation, selection of targets  for the upcoming
runs, using  custom software  during observations  (including pointing
the telescope), data processing pipeline, and archiving the results in
the custom database which counts 46,113 entries as of January 1, 2026.
The latest  results are  reflected in the  program, allowing  to plan
next observations.  This closed-loop process with a fast duty cycle is
particularly   suitable   for   monitoring   fast   orbital   motions.
Determination  of orbits  is  a time-domain  science  that requires  
flexible  and regular  access to  the telescope  rather than  standard
scheduling per semester.

\subsection{Instrument}
\label{sec:inst}

The instrument and  observing procedure are described  in the previous
papers of these series \citep[e.g.][]{SAM19} and briefly summarized by
\citet{SAM22}. 
 We used  mostly the  near-infrared  $I$ filter  (824/170\,nm), while  the
Str\"omgren  $y$ filter  (543/22\,nm) was  chosen for  brighter and/or
closer pairs. The pixel size  is 15.3\,mas.  In the standard observing
mode, two  data cubes of  200$\times$200 pixels (3\farcs15  field) and
400--600 frames each are taken on each target with an exposure time of
25\,ms.   A single  reference star  is observed  immediately after  or
before,  if needed.   Pairs wider  that 1\farcs5  are observed  with a
400$\times$400 format and, optionally, with a 2$\times$2 binning. Each
data  cube  is used  to  compute  the  power spectrum,  the  high-pass
filtered speckle autocorrelation function  (ACF), the averge centered
and the shift-and-add images.

The HRCam normally receives the light through the SOAR adaptive optics
module, SAM  \citep{SAM}. The  SAM was removed  from the  telescope in
January 2024 for upgrade.  The HRCam was installed without SAM using a
modified mechanical  adaptor and tested  on the engineering  nights of
2024 January  26 and February  26-27.  The atmospheric  dispersion was
compensated  by  the  regular SOAR  atmospheric  dispersion  corrector
(ADC), instead  of the SAM  ADC.  These tests demonstrated  that HRCam
can work without SAM, if needed, without affecting its performance and
results. The new  speckle instrument HRCam2 based on  a CMOS camera
will be  installed at the  asquisition port  and will employ  the SOAR
ADC. HRCam2 has been tested on sky in 2026 January. 

Meanwhile,  the  SAM   was  upgraded  in 2024 February  to  SAMplus
\citep{SAMplus},  including, among  other things,  replacement of  its
deformable mirror  (DM).  SAMplus  was tested  during the  March 25--26
engineering run,  using HRCam as an  imager. From that date  on, HRCam
again receives  the light through  SAM with  its upgraded DM,  and the
regular SAM  ADC is used.  The  first scheduled science night  in this
configuration was 2024 March 28.  Differences in the instrument
  configuration do not affect such HRCam characteristics as
  sensitivity, resolution, and dynamic range. 

The diffraction limit  $\lambda/D$ of the 4.1 m SOAR  telescope in the
$y$ and  $I$ filters is  27 and  41 mas, respectively.   However, many
pairs  at  closer  separations could be measured.   Their  positions  are
obtained by  modeling the speckle power  spectra of the target  and of
the reference star; they are less accurate and are marked by colons.
Typical HRCam detection limits (contrast vs. separation) are
illustrated in Figure~1 of \citet{SAM23},  see also
  Section~\ref{sec:TESS}. 
 The  SOAR  speckle  instrument
  usually  gets data  on  300 targets  per  night, surpassing  typical
  efficiency of speckle instruments at other telescopes by a factor of
  $\sim$5.

\subsection{Image Quality}
\label{sec:psf}

The signal to  noise ratio in speckle  interferometry strongly depends
on the  object flux and  seeing.  This  relation has been  explored by
\citet{HRCam2024} using both the real  HRCam data and simulations.  In
2024--2025, the  seeing at Cerro  Tololo and Cerro Pach\'on  was worse
than usual.  The median Full Width  at Half Maximum (FWHM) of centered
images  determined  by the  speckle  pipeline  is 0\farcs954  for  all
 2024-2025 data.  Among the  21 observing  runs,  the smallest  and largest  FWHM
medians  are  0\farcs83  and   1\farcs35,  respectively.  Poor  seeing
increases the measurement errors for faint stars.  Some observing runs
were affected by thin clouds (cirrus).

\begin{figure}[ht]
\epsscale{1.1}
\plotone{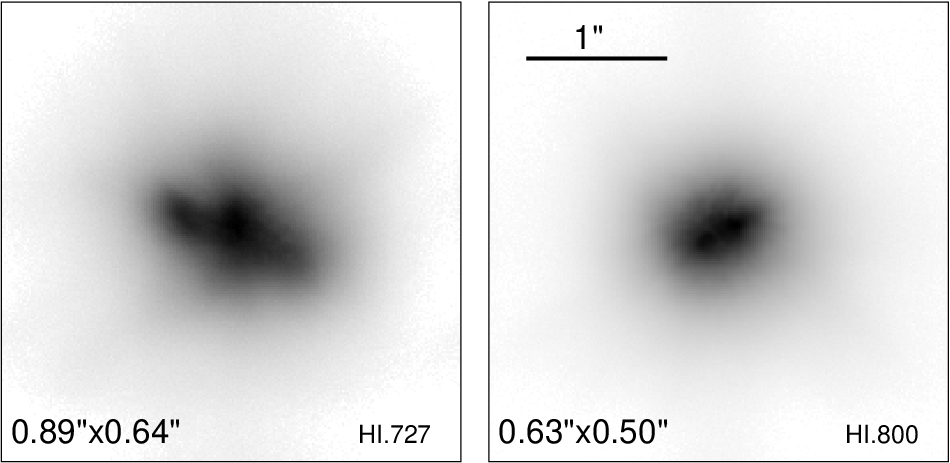}
\caption{Mean centered images of single stars (in negative square-root
  rendering)  recorded  with  HRCam   on  2024  November  16/17  under
  excellent seeing at  an elevation around 60\degr ~in  the $I$ filter.
  The size  of each image is  3\farcs15. The FWHM along  horizontal and
  vertical directions  is indicated. The first  image HI.727, recorded
  immediately  after tuning  the  SOAR optics,  shows  clear signs  of
  residual  aberrations.  The  image  HI.800 taken  an  hour later  is
  sharper, although still somewhat asymmetric.
\label{fig:PSF} }
\end{figure}

The size  of the stellar images  at SOAR is enlarged,  compared to the
seeing, by  two effects: inhomogeneous  air in the dome  (dome seeing)
and optical  aberrations.  The dome seeing  causes characteristic slow
and irregular distortions of the  images dominated by random low-order
aberrations  like  defocus  or  astigmatism;  they  are  qualitatively
different from the normal (external) seeing.  Although the dome-seeing
distortions are obvious in some speckle movies, presently they are not
quantified.   The  optical  aberrations,  on the  other  hand,  become
apparent under very good seeing.  The night of 2024 November 16/17 had
excellent  conditions:  the  seeing  monitor  indicated  values  below
0\farcs5  for most  of the  night, and  the wind  was slow.   Yet, the
median  FWHM  of  the  speckle  images  in  this  run  was  0\farcs83.
Re-tuning the  SOAR optics  in the  middle of the  night did  not help
much.   The aberrations  are revealed  by the  irregular shape  of the
average centered image derived from the data cube.  For example, the
image  cube HI.727  taken  immediately after  tuning  the SOAR  optics
yields  a   FWHM  size   of  0\farcs89$   \times$0\farcs64  (elongated
horizontally) and  a spotty structure,  while the external  seeing was
better  than 0\farcs4  (Figure~\ref{fig:PSF}, left).   The aberrations
evolve  with  time  and  telescope elevation.   Occasionally,  a  FWHM
resolution  of 0\farcs5  is attained  in  one or  both directions,  as
illustrated  in Figure~\ref{fig:PSF}, right.   Aberrations under  good seeing
cause systematic distortions of the  speckle power spectra, hence they
degrade  the  sensitivity  and  the astrometric  accuracy  of  speckle
interferometry.  The observer should control the telescope focus while
keeping an eye on evolving conditions and image quality.

On  2024  February  26/27  and  27/28,  the  HRCam  observations  were
adversely  affected by  the  jitter  of the  SOAR  mount.  The  jitter
appeared   and  disappeared   intermittently,  usually   at  telescope
elevation of 50\degr and below.  The oscillations were quasi-periodic,
with the dominant frequency between 1\,Hz and 2.5\,Hz.  The rms motion
of the star centroid without jitter is typically about 0\farcs3 in a
10 s data  cube.  The oscillations increase the  rms to $\sim$1\arcsec
or more.  Motion of the star  in one direction smears the speckles and
strongly  distorts the  speckle  power  spectrum.  The  non-stationary
nature  of the  oscillations prevents  calibration of  this distortion
by reference  stars.  In one  case, the oscillation was  present in
one data  cube and stopped in  the following cube taken  a few seconds
later on the same target.  Similar oscillations with smaller amplitude
were encountered in some other  runs. Periodic errors of the encoders,
partially amplified by the response of  the mount servo, is one of the
known causes of oscillations; hardware problems in the telescope mount
drives is another.

\subsection{Updated Calibration and Astrometric Accuracy of HRCam}
\label{sec:cal}

\begin{figure}[ht]
\epsscale{1.1}
\plotone{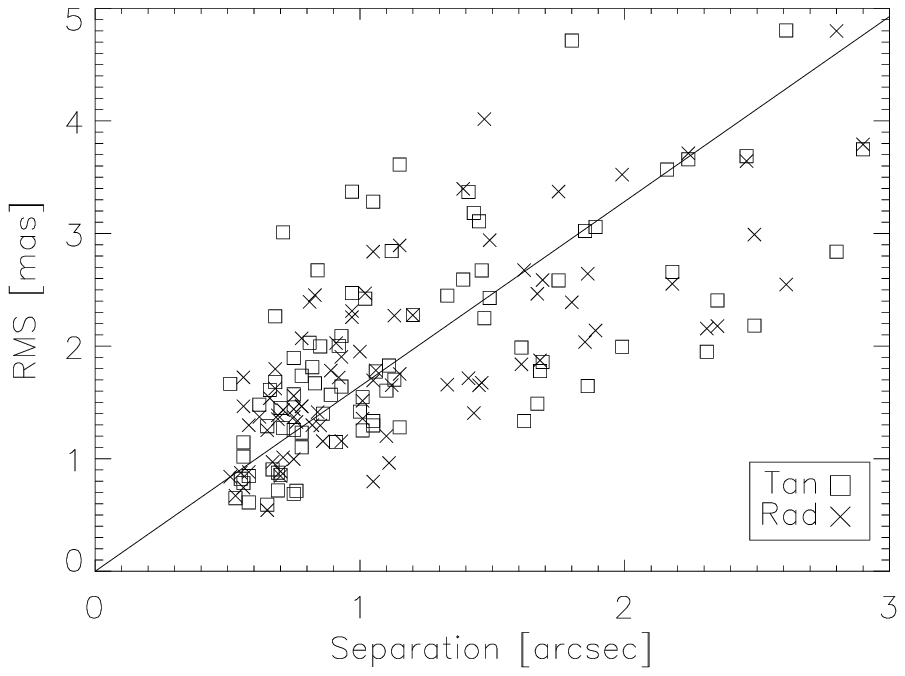}
\caption{The rms deviations of the calibrator binaries from their models
  in the tangential (squares) and radial (crosses) direction vs. 
  angular separation. The line is a linear approximation $\sigma
  \approx 1.64 \rho$  which corresponds to the relative pixel
    scale accuracy of 0.16\% or the position angle accuracy of 0\fdg09. 
\label{fig:cal} }
\end{figure}

Calibration of  the pixel scale and  orientation of HRCam is  based on
observations  of relatively  wide pairs, called  {\em calibrators}
for brevity.   Although the orbits of  many ``slow'' pairs are  of low
quality or unknown, the small observed arcs can be accurately modeled.
As the  calibrator models improve with  time, the accuracy of  the old
(published) data can be improved as well by retroactive calibration.

The  set of  calibrators  and their  models  were revised  iteratively
several times.  The  previous list of 104 calibrators used  the data up
to 2021.75 \citep{SAM21}.  These models  were checked against Gaia data
release 3  (GDR3) \citep{Gaia3}, and  a minor systematic error  in the
overall  calibration  was  corrected  from   2021  on.   In  2025  the
calibrator set was revised again  and reduced to 86 pairs, eliminating
two  pairs  (J01316$-$5322  and  J18250$-$0135)  with  inner  subsystems
detected  via  wobble  \citep{TRI26}   and  several  calibrators  with
magnitude differences above 3 mag. 

The  calibrator models  (73 orbits  and 13  linear) were  updated here
using the SOAR  data till 2025.77 (see Appendix).   The rms deviations
$\sigma$ between the  data (relative positions) and the  models in the
tangential  and  radial  directions  are similar:  their  medians  are
1.78\,mas and  1.72\,mas, respectively.  The deviation  increases with
the separation  $\rho$ (in arcseconds)  as $\sigma \approx  1.64 \rho$
mas, corresponding  to an accuracy  of 0.16\%  in the pixel  scale and
0\fdg09 in  position angle (Figure~\ref{fig:cal}).   Comparison of
the  updated  models  with   GDR3  relative  positions  reveals  no
measurable  systematic  differences.   The   GDR3  astrometry  of  both
components  is  available for  48  calibrators  (the rest  are  either
unresolved by Gaia or lack  5-parameter solutions for some components,
making  their  GDR3  relative  positions  unreliable).   For  this
subset, the mean difference between the calibrator models interpolated
to  2016.0 and  the  GDR3 positions  in the  relative  pixel scale  is
1.2\,10$^{-5}$; the  rms difference of 0.12\% in  scale is similar
to the internal  accuracy of our calibrator models,  0.16\%.  The mean
difference  in position  angle between  SOAR calibrators  and GDR3  is
+0\fdg022, the rms scatter is 0\fdg036,  and the rms difference in the
tangential direction is 1.77\,mas.

Using the revised  calibrator models, we can correct  all SOAR speckle
data  retroactively  to improve  their  accuracy.   The data  of  some
observing runs  are less  accurate for  instrumental reasons,  e.g.  a
detector  with  poor  charge  transfer  efficiency  used  in  2014.75,
2014.86, and 2016.95, or problems  of the SOAR Nasmyth rotator control
in 2018.25.  For some calibrators,  a few deviant points  were removed
while fitting  the models. We  plan to  publish all SOAR  speckle data
with these corrections in the future.

The  published HRCam  results  contain only  the internal  measurement
errors $\sigma_i$  in the tangential  and radial directions.   For the
data of 2024--25, the median  estimated internal errors $\sigma_i$ are
0.42  and   0.39  mas  in   the  radial  and   tangential  directions,
respectively  (for  comparison, the median internal  errors of the
  HRCam measures  in 2023 are  0.32\,mas).  The  external measurement
error $\sigma_e$ can  be estimated as a quadratic sum  of the internal
error and the calibration uncertainty:
\begin{equation}
\sigma^2_{e} = \sigma^2_i + (C \rho)^2 ,
\label{eq:ext}
\end{equation}
where $C =  1.6$\,mas/arcsec.  For our data, the  median separation of
0\farcs19 corresponds to the calibration  error $C \rho$ of 0.30\,mas,
so for most close pairs the calibration is only a minor contributor to
the astrometric errors.  In fitting  the orbits, measurement errors of
2\,mas are  usually adopted,  in agreement  with Figure~\ref{fig:cal};
for close and  bright pairs with reliable orbits the  residuals of the
SOAR measures are often below 1\,mas.

\section{Results}
\label{sec:res}

\subsection{Data Tables}
\label{sec:tables}

\begin{figure}[ht]
\epsscale{1.1}
\plotone{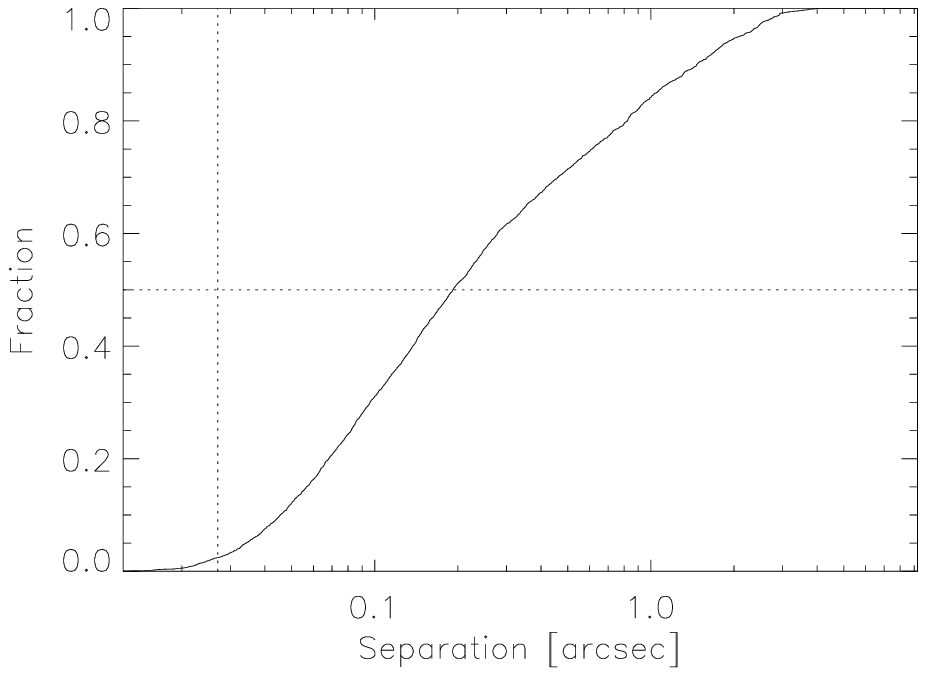}
\caption{Cumulative   distribution    of   angular    separations   in
  Table~\ref{tab:measures}.  The   vertical  dotted  line   shows  the
  diffraction limit of 27\,mas in the $y$ filter.
\label{fig:cumhist} }
\end{figure}

The results (measures of resolved pairs and non-resolutions) are presented in 
the same format as in \citet{SAM23}. The long tables are published
electronically; here we describe their content.

In both tables, each object is identified by its WDS code based on the
J2000  position (such  codes  are  generated if  missing  in the  WDS,
e.g. for  single stars), discovery designation  (DD), alternative name
(when  available), and  the  J2000  coordinates. The  DD  is a  name
assigned by the WDS compilers to each pair. As noted in \citet{SAM23},
the DDs  are obsolete and   nowadays they are used infrequently  by the
astronomical community; we  keep them for compatibility  with the WDS.
For pairs found only in the WDS supplement, WDSS,\footnote{
  \url{http://www.astro.gsu.edu/wds/Supplement/wdss_summ.txt}} the 
coordinate-based WDSS 14-character codes are listed instead of the DDs. 


\begin{deluxetable}{ l l l l }
\tabletypesize{\scriptsize}
\tablewidth{0pt}
\tablecaption{Measurements of Double Stars at SOAR 
\label{tab:measures}}
\tablehead{
\colhead{Col.} &
\colhead{Label} &
\colhead{Format} &
\colhead{Description, units} 
}
\startdata
\phn1 & WDS                  & A10  & WDS code (J2000)  \\
\phn2 & Discov.              & A16  & Discoverer Designation   \\
\phn3 & Other                & A16  & Alternative name \\
\phn4 & R.A.                 & F9.5 & R.A.\ J2000 (deg) \\
\phn5 & Decl.                & F9.5 & Declination J2000 (deg) \\
\phn6 & Epoch                & F9.4 & Julian year (yr) \\
\phn7 & Filt.                & A2   & Filter \\
\phn8 & $N$                  & I2   & Number of averaged cubes \\
\phn9 & $\theta$             & F8.1 & Position angle (deg) \\
   10 & $\rho \sigma_\theta$ & F5.1 & Tangential error (mas) \\
   11 & $\rho$               & F8.4 & Separation (arcsec) \\
   12 & $\sigma_\rho$        & F5.1 & Radial error (mas) \\
   13 & $\Delta m$           & F7.1 & Magnitude difference (mag) \\
   14 & Flag                 & A1   & Flag of magnitude difference\tablenotemark{a} \\
   15 & Tag                 & A1   & System tag\tablenotemark{b} \\
   16 & (O$-$C)$_\theta$     & F8.1 & Residual in angle (deg) \\
   17 & (O$-$C)$_\rho$       & F8.3 & Residual in separation (arcsec) \\
   18 & Ref                  & A9   & Orbit reference\tablenotemark{c} 
\enddata
\tablenotetext{a}{Magnitude flags: 
q -- the quadrant is determined; 
* -- $\Delta m$ and quadrant from average image; 
: -- noisy data or tentative measure. }
\tablenotetext{b}{System tags: 
A -- Hipparcos-Gaia acceleration stars (K.~Franson);
C -- Wide pairs observed for J.~Chanam\'e;
G -- Wide pairs with relative positions in GDR3, likely physical;
g -- Wide pairs with relative positions in GDR3, likely optical;
M -- M-type dwarfs within 30\,pc \citep{Vrijmoet2022,Vrijmoet2026};
N -- New pair resolved in 2024--2025;
P -- Hierarchical systems from \citet{Powell2023}; 
T -- Hierarchies within 100 pc \citep{GKM};
Z -- TESS objects of interest \citep{TESS2}.
}
\tablenotetext{c}{Orbit References are provided at
  \url{https://crf.usno.navy.mil/data_products/WDS/orb6/wdsref.html} }
(This table is available in its entirety in machine-readable form in the online article.)
\end{deluxetable}

\begin{deluxetable}{ l l l l }
\tabletypesize{\scriptsize}
\tablewidth{0pt}
\tablecaption{Unresolved Stars 
\label{tab:single}}
\tablehead{
\colhead{Col.} &
\colhead{Label} &
\colhead{Format} &
\colhead{Description, units} 
}
\startdata
\phn1 & WDS              & A10  & WDS code (J2000)  \\
\phn2 & Discov.\         & A16  & Discoverer Designation  \\
\phn3 & Other            & A16  & Alternative name \\
\phn4 & R.A.            & F9.5 & R.A. J2000 (deg) \\
\phn5 & Decl.            & F9.5 & Declination J2000 (deg) \\
\phn6 & Epoch            & F9.4 & Julian year  (yr) \\
\phn7 & Filt.\           & A2   & Filter \\
\phn8 & $N$              & I2   & Number of averaged cubes \\
\phn9 & $\rho_{\rm min}$ & F7.3 & Angular resolution (arcsec)  \\
   10 & $\Delta m$(0.15) & F7.2 & Max. $\Delta m$ at 0\farcs15 (mag) \\
   11 & $\Delta m$(1)    & F7.2 & Max. $\Delta m$ at 1\arcsec (mag) \\
\enddata
(This table is available in its entirety in machine-readable form in the online article.)
\end{deluxetable}

Table~\ref{tab:measures}  lists  5316 measures  of  3532
resolved   pairs    and   subsystems,   including    new   discoveries
(Figure~\ref{fig:cumhist}).   Equatorial  coordinates  for  the  epoch
J2000  in degrees  are  given in  columns (4)  and  (5) to  facilitate
matching  with  other  catalogs.   Circumstances  of  this  particular
observation (Julian  year, filter,  number of  averaged cubes),  be it
Table 1  or 2, are given  in columns (6)  through (8). In the  case of
resolved  multiple  systems,  the positional  measurements  and  their
errors (columns 9--12) and magnitude  differences (column 13) refer to
the   individual   pairings   between   components,   not   to   their
photocenters. As  in the previous papers  of this series, we  list the
internal errors derived  from the speckle power  spectrum modeling and
from the difference between the measures obtained from two data cubes.
The real (external) errors are larger (see equation~\ref{eq:ext}).

The flags in column (14) indicate the cases where the true quadrant is
determined (otherwise the position angle is measured modulo 180\degr),
when  the  relative photometry  of  wide  pairs  is derived  from  the
long-exposure  images (this  reduces  the $\Delta  m$  bias caused  by
speckle  anisoplanatism),  and   when  the  data  are   noisy  or  the
resolutions are tentative (see TMH10).

To facilitate identification  of pairs that either  have been resolved
previously but remain unpublished or are published but not yet entered
in the WDS,  we provide in column (15) one-character  tags.  Their
  meaning and the numbers of resolved individual systems with  each tag are as follows:  A
--- acceleration  stars (K.~Franson,  B.~Bowler, $N=51$), C  -- subsystems  in
wide  pairs  (J.~Chanam\'e, $N=16$), G  or  g  ---  wide pairs  with  relative
positions     in     GDR3 (121 and 78, respectively),     M    --     nearby     M-type     stars
\citep[][$N=8$]{Vrijmoet2022,Vrijmoet2026}, N  --- new pairs resolved  here for
the  first time  (Section~\ref{sec:new}, $N=353$),  P ---  hierarchical systems
published  by  \citet[][$N=10$]{Powell2023},  T  ---  components  of  triple  or
higher-order  hierarchies  within  100\,pc  \citep[][$N=18$]{GKM},  Z  ---  TESS
objects of interest \citep[][$N=2$]{TESS2}.

For binary stars with known orbits,  the residuals to the latest orbit
(see Section~\ref{sec:orbits}) and its  reference are provided in columns
(16)--(18). Residuals close to 180\degr ~mean that the orbit swaps the
brighter  (A) and  fainter (B)  stars.  However, in  some binaries  or
triples the secondary is fainter in  one filter and brighter in other.
In these cases, it is better  to keep the historical identification of
the components  in agreement  with the  orbit and  to assign  a negative
magnitude difference $\Delta m$.

The  non-resolutions  of  1723 targets (mostly  reference  stars)  are
reported in Table~\ref{tab:single}.  Its first columns (1) to (8) have
the same meaning and format as in Table~\ref{tab:measures}. Column (9)
gives the minimum resolvable separation when pairs with $\Delta m < 1$
mag are detectable. It is  computed from the maximum spatial frequency
of the  useful signal in the  power spectrum and is  normally close to
the  formal  diffraction  limit $\lambda/D$.  The  following
columns (10) and (11) provide  the indicative dynamic range, i.e., the
maximum magnitude difference at separations of 0\farcs15 and 1\arcsec,
respectively, at $5\sigma$ detection level.

\subsection{Nearby Hierarchical Systems}
\label{sec:mult}

\begin{figure}[ht]
\epsscale{1.1}
\plotone{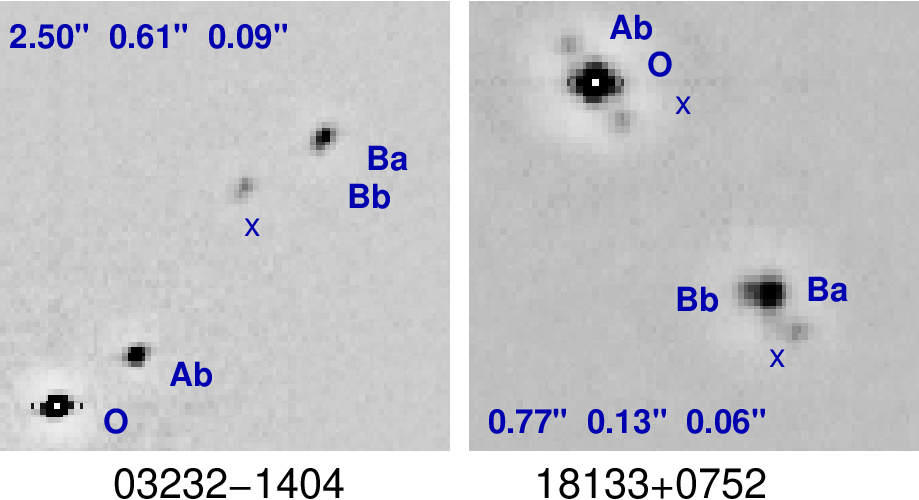}
\caption{Fragments  of speckle  ACFs  of two  new  2+2 quadruples,  in
  negative  rendering. The  orientation  is standard  (north up,  east
  left). The  components corresponding  to specific peaks  are marked,
  with  the  ACF  center  marked  by  O and  by  a  white  dot.  Peaks
  corresponding to cross-correlations between secondaries, e.g. Ba,Ab,
  are marked  by 'x'  (ACF of  a resolved  quadruple has  12 secondary
  peaks).  The outer and two inner separations are indicated.
\label{fig:quad} }
\end{figure}

\begin{figure}[ht]
\epsscale{1.1}
\plotone{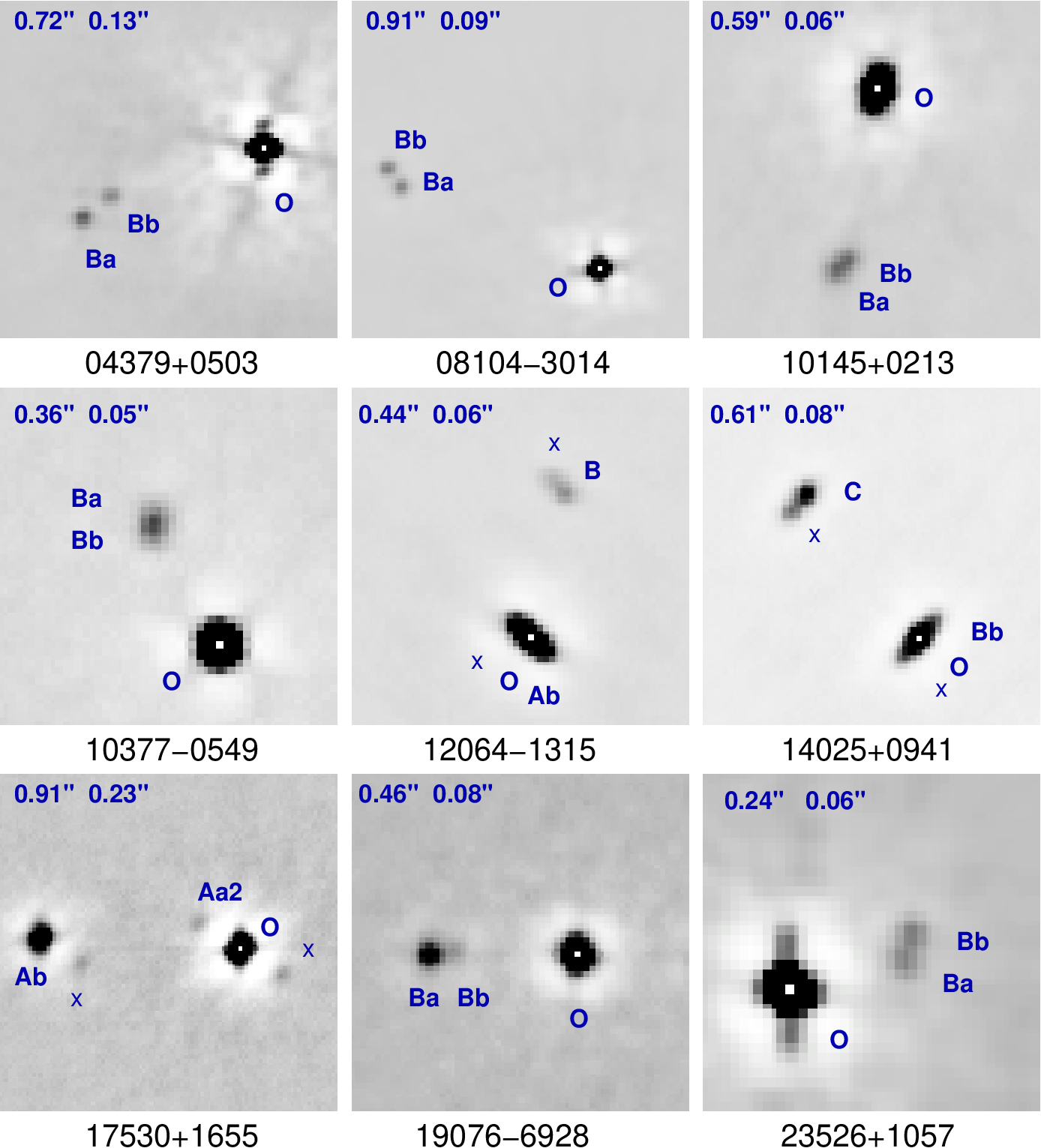}
\caption{Speckle ACFs of 9 new close triples with sub-arcsecond outer
  separations and separation ratios above 1:10. 
\label{fig:close} 
}
\end{figure}

Monitoring orbital motions in nearby  resolved multiple systems is the
core of  the SOAR speckle  program.  Accumulation of the  data defines
the inner and  outer orbits, thus constraining the  dynamical state of
these hierarchical systems. Statistics  of mutual inclinations, period
ratios,  and masses  can be  compared  to the  predictions of  various
formation scenarios. It is already clear that hierarchies in the field
originate from  different formation channels, but  their relative role
and its  dependence on  the environment and  stellar mass  are debated
\citep{review}.  Recent observational  work based on the  SOAR data is
published  in  \citep{TRI23,TRI25,TRI26,Tok2026}.    For  some  stars,
speckle   interferometry   is    combined   with   radial   velocities
\citep{CHI23,CHI25}.    Speckle  interferometry   was  also   used  by
\citet{Hartmann2025}  to resolve  inner subsystems  in wide  binaries.
\citet{Mendez2025} characterized several  compact hierarchies with the
Gemini  8  m  telescope.   Very   tight  subsystems  are  resolved  by
interferometers such as CHARA \citep[e.g.][]{Torres2022}.

The SOAR angular resolution, 0\farcs03, together with the third Kepler
law, mean  that periods  short enough  to be  measurable on  the human
lifetime scale are  limited to relatively nearby stars:  at a distance
of 100\,pc,  the resolution limit  of 3 au corresponds  to decade-long
periods, while wider pairs move  more slowly. So, the most interesting
targets are nearby binaries and  triples.  Our attention is focused on
stars within  100 pc.   Over a thousand  new candidate  hierarchies in
this volume were identified in the  Gaia catalog of nearby stars, GCNS
\citep{GCNS} as  source pairs with  common distance and  proper motion
(PM) where at least one star had  signs of inner subsystems such as an
increased  reduced unit  weight error  (RUWE) or  multi-peak transits;
$\sim$500 inner subsystems in these  hierarchies were resolved at SOAR
in 2022--2023  \citep{GKM}.  Periods  of the closest  subsystems range
from  years  to  decades,  and  their fast  orbital  motion  is  being
monitored at SOAR.  In 2024--2025,  we measured 357 such targets.  One
of  those, WDS  J13038$-$3645,  is  not confirmed  and  appears to  be
spurious.

Many interesting hierarchies have been discovered at SOAR accidentally
when previously known binaries from the WDS in need of modern measures
were targeted.   This led us  to the idea  of observing all  WDS pairs
within 100 pc that lack previous SOAR data, in hope of discovering new
hierarchies. Note  that many binaries  and triples with  components of
comparable  masses are  missing from  the  GCNS because  Gaia did  not
determine their parallaxes. Only  systems with substantial contrast or
very small  separations where  single-star astrometric  solutions were
accepted by Gaia are considered here.  We selected from the GCNS stars
brighter than $G=13$ mag south  of declination +20\degr whose position
matches WDS pairs with separations under 3\arcsec.  These criteria are
dictated by the capabilities of HRCam.  In this sample, there are 1927
WDS pairs  that lack any HRCam  data.  A subset of  401 most promising
pairs from this list with RUWE above 1.4 was included in the observing
program, and  375 of those  were observed as ``fillers''  in the
general speckle program.

\begin{deluxetable}{l l  c  c c l } 
\tabletypesize{\scriptsize}
\tablewidth{0pt}
\tablecaption{New Resolved Triples Within 100\,pc
\label{tab:trip}}
\tablehead{
\colhead{WDS} &
\colhead{Comp.} &
\colhead{$\rho_{\rm out}$ } &
\colhead{$\rho_{\rm in}$} &
\colhead{$\Delta m_{\rm in}$ } &
\colhead{DD} \\
&        &
\colhead{(\arcsec)} &
\colhead{(\arcsec)} &
\colhead{(mag)}  &
}
\startdata
00049$-$3230 & Aa,B  & 1.32 & 0.060 & 0.3   &  SEE 500 \\ 
00581$-$5742 & Aa,B  & 0.89 & 0.081 & 2.6   & B 1027 \\
02195$-$3137 & Aa,B  & 1.70 & 0.095 & 2.4   & KPP2848 \\
03232$-$1404 &Aa1,Ab & 2.50 & 0.614 & 1.0   & DAM1303 \\
03232$-$1404 &Aa,Ab1 & 2.50 & 0.085 & 0.6   & DAM1303 \\
03552+0417   & A,Ba  & 1.36 & 0.149 & 0.1   & A 2349 \\
04173$+$1214 & Aa,B  & 0.71 & 0.049 & 0.4   & RAO 544 \\
04379$+$0503 & A,Ba  & 0.73 & 0.135 & 0.4   & ELP 10 \\ 
05089$-$1255 & A,Ba  & 1.50 & 0.154 & 0.2   & RST3413 \\
05130$-$7028 & Aa,B  & 1.69 & 0.079 & 0.0   & JNN 32 \\
05496$-$1429 & A,Ba  & 2.31 & 0.141 & 1.9   & BU 94 \\
07494$-$3033 & A,Ba  & 1.53 & 0.237 & 3.6   & I 186  \\
08104$-$3014 & A,Ba  & 0.91 & 0.094 & 0.0   & I 791 \\
09002$+$1550 & A,Ba  & 1.93 & 0.163 & 2.9   & ALD 115 \\
10145+0213   & A,Ba  & 0.59 & 0.059 & 0.1   & RAO 572 \\
10377$-$0549 & A,Ba  & 0.36 & 0.051 & 0.0   & JNN 73 \\ 
12064$-$1315 & Aa,B  & 0.44 & 0.058 & 0.5   & JNN 77 \\
12153$-$3204 &Aa1,Ab & 1.53 & 0.076 & 0.6   & LDS4197 \\
12203$-$5242 & Aa,B  & 2.58 & 0.149 & 1.7   & BRT2072 \\
14025$+$0941 &Ba,C   & 0.61 & 0.078 & 0.6   & RAO 309 \\
14306$+$0306 & A,Ba  & 2.56 & 0.433 & 0.0   & RAO 588 \\
14562+1745   & A,Ca  & 1.90 & 0.047 & 0.1   & GII 61 \\ 
16254$-$2710 & A,Ba  & 2.92 & 0.335 & 1.9   &  LDS4666 \\
16270$-$6944 & A,Ba  & 1.97 & 0.221 & 0.52  & NSN 351 \\ 
16535$-$6049 & Aa,B  & 1.81 & 0.059 & 1.4   &  B 2396 \\ 
17080$-$1929 & Aa,B  & 2.47 & 0.040 & 0.7   & NSN 139 \\
17530$+$1655 &Aa1,Ab & 0.91 & 0.227 & 3.3   &  CRC 27 \\ 
18133$+$0752 &Aa,B   & 0.77 & 0.128 & 1.2   &  RAO 606 \\ 
18133$+$0752 & A,Ba  & 0.77 & 0.062 & 0.9   &  RAO 606 \\ 
18574$-$3352 & A,BC  & 2.39 & 0.330 & 0.8   &  B 956 \\
19076$-$6928 &A,Ba   & 0.46 & 0.083 & 1.7   &  TDT1240 \\ 
19252$+$0227 &A,Ba   & 1.95 & 0.106 & 3.1   &  STF2513 \\ 
19344$-$7657 &Aa,B   & 2.19 & 0.228 & 0.8   &  KPP4155 \\
19475$-$2150 &A,Ba   & 1.37 & 0.133 & 2.6   &  STN 49 \\
19540$+$1518 &Aa,B   & 2.22 & 0.090 & 3.0   &  STF2596 \\ 
19552$-$0051 &Aa,B   & 1.96 & 0.108 & 0.1   &  BU 830 \\ 
23097$-$0158 &Aa,B   & 1.66 & 0.045 & 0.6   &  CRC 76 \\ 
23414$-$0838 &A,Ba   & 1.75 & 0.130 & 2.0   &  A 423 \\ 
23526$+$1057 &A,Ba   & 0.24 & 0.058 & 0.1   & YSC 17 \\
\enddata
\end{deluxetable}

The first  results of  this effort are  encouraging.  Thirty  five new
triples and two quadruples were  discovered (a 10\% ``success rate'');
they are listed in Table~\ref{tab:trip}.  Some of those were confirmed
by subsequent SOAR observations. The  new inner pairs were resolved in
20  secondary   and  19  primary   components.   Figure~\ref{fig:quad}
illustrates the two new quadruples, while Figure~\ref{fig:close} shows
nine  new compact  triples with  comparable separations  identified in
this project.  It is noteworthy  that some outer pairs were discovered
in the 21st  century using adaptive optics  or speckle interferometry,
but their inner  subsystems have been missed.   Of particular interest
are so-called double twins, were  faint secondaries are actually pairs
of  similar low-mass  stars.  Such  hierarchies often  have comparable
inner    and   outer    separations    and   approximately    coplanar
low-eccentricity orbits,  like the  emblematic low-mass  triple system
LHS~1070  \citep{DancingTwins}.   Many  inner subsystems  in  the  new
hierarchies  have short  estimated periods,  and their  orbits can  be
determined (if the monitoring at SOAR continues) to probe the internal
dynamics. In WDS J19540+1518 (STF 2596), orbit of the inner pair Aa,Ab
with a  period of 68  yr was derived from  the wobble in  the archival
measures; the faint star Ab turned out  to be as massive as Aa and was
revealed as an eclipsing pair; so,  this classical visual binary is in
fact a quadruple system \citep{TRI26}.


The  majority of  candidate triples  identified by  the elevated  RUWE
remained unresolved.  Some  of those are real triples  where the inner
companions are too  close and/or too faint for  speckle resolution. In
some other cases, the RUWE could be increased by the orbital motion in
the wide pairs or by occasional blending of their components in Gaia.
Close nearby  pairs have an  elevated RUWE caused by  their relatively
fast orbital motion. Measurements at  SOAR, combined with the existing
data, can define their orbits.  Several first-time orbits were derived
from single SOAR  measures of neglected pairs within  100 pc. However,
most  these  orbits are  preliminary  owing  to scarce  or  inaccurate
historic data.

\subsection{New Metal-poor Binaries}
\label{sec:metal-poor}

In 2025 we started monitoring nearby metal-poor stars with the purpose
of creating  a sizable sample  of local Galactic halo  binaries.  This
effort  extends  our  long-term  program  on  securing  data  for  the
mass-luminosity-metallicity   relation,  MLR   \citep{Mendez2021}.   A
comparison of empirical masses derived from visual orbits with stellar
evolutionary models  can be found in  Figure~10 of \citet{Horch2015a};
see  also an  update in  \citet{Horch2019}.  Their  analysis, however,
compares masses with colors which  are affected by 
 line-blanketing (selective absorption by spectral lines). 
We plan to use luminosities,  as in \citet{Mann2019}.  Our calculation
of the theoretical MLR using  MESA isochrones indicates that for these
objects,  near-IR  photometry  is rather  insensitive  to  metallicity
effects,  a fact  that \citet{Mann2019}  indeed point  out (see  their
Figure 16). Instead, optical photometry should be used, also extending
the metalliciy  range to  ${\rm [Fe/H]} <  -0.5$. We  furthermore note
that the  MLR analysis  should be  restricted to  objects on  the main
sequence, which  for the age of  the Halo implies masses  of less than
about 0.85 \msun.

In order to  have the best chance to resolve  metal-poor stars, we constrained our
sample  to  objects  closer  than  100  pc,  selected  from  the  GCNS
\citep{GCNS}.  This catalog, which is based on GDR3, does not give
metallicities  directly,  but a  cross  match  with the  general  GDR3
catalog does provide such information.   The metallicities in GDR3 are
determined from the on-board Blue  (BP) and Red (RP) prism photometers
which collect low resolution spectrophotometric measurements of 
spectral energy  distributions over the wavelength  ranges 330–-680 nm
and  630--1050   nm,  respectively;  and  from   the  radial  velocity
spectrometer (RVS)  which collects medium resolution  ($R \sim 11700$)
spectra over the wavelength range  845--872 nm centered on the Calcium
triplet   region   \citep{Gaia1,   Recino2023,   Andrae2023}.    These
metallicities  are  limited  to   objects  brighter  than  $G=12$  mag
\citep{Katz2023}, which is a good  match to the HRCam magnitude limit.
Eventually   one   could   also  crossmatch   with   other   suitable
spectroscopic  catalogs to  further enlarge  the sample  with measured
metallicities.

The GCNS has  over 330,000 entries. For a local  halo normalization of
1\%, one would  expect to have over 3000 local  subdwarfs.  If roughly
half  of them  are  in  binary systems   \citep[as  for the  disk
    stars,][]{Raghavan2010}   then the   sample of  Galactic halo
  binaries   would   be  around   1500   pairs   over  the   whole
sky. 
We can improve
the binary  detection rate by  observing only those objects  that have
some indication of binarity in the Gaia catalog, particularly the RUWE
\citep{Lindegren2021},    and    the   {\tt    ipd\_frac\_multi\_peak}
\citep{Penoyre2022, Gaia3}, and also  the F2 goodness-of-fit parameter
in         the         Hipparcos         catalog         \footnote{See
  \url{https://hipparcos-tools.cosmos.esa.int/pstex/sect2_01.pdf}.}. Based
on  these constrains,  we selected  low-metallicity binary  candidates
([Fe/H] $ < -1.2$  dex) with $ V < 12$ or $G  < 11.5$ mag and negative
declination.  After  discarding already  known binaries listed  in the
WDS, we ended up with 87 targets,  all of which were observed.  Out of
these, 43 pairs (including triples)  were resolved (tag N in Table~1),
with another  8 wider  pairs also found  in GDR3 (tags  G or  g).  The
selection of binary candidates using GDR3 is therefore very efficient,
leading to a ``success rate'' over 50\%.

The closest new  pairs are expected to move fast,  so their orbits can
be  determined  within  several   years,  leading  eventually  to  the
measurement  of   masses.   For  example,  the   estimated  period  of
J02270$-$0326  is  $\sim$7  yr.     Interestingly, five subdwarfs are
revealed as triples: J01483$-$7004  (Aab-B, 0\farcs065 and 0\farcs51),
J01516$-$3757 (0\farcs45 and 0\farcs71, trapezium-like), J03258$-$1225
(Aab-B, 0\farcs046 and 1\arcsec),  J14376$-$6021 (A-Bab, 0\farcs94 and
0\farcs051), and J19246$-$0839 (A-Bab, 0\farcs38 and 0\farcs044).

Of course,  the newly  detected binaries have  to be  confirmed through
subsequent observations  to discard optical pairs,  which are unlikely
given  the  small  separations,  and to  start  accumulating  data  to
constrain  their orbits.   Eventually  we would  like  to extend  this
sample by including stars  with intermediate (thick-disk) metallicity,
$-1.0 < {\rm [Fe/H]} < -0.4$ dex.

\subsection{TESS Objects of Interest}
\label{sec:TESS}

Since 2018, HRCam has been used to detect close companions to the TESS
objects  of   interest  (TOIs),  mostly  candidate   exohosts.   These
observations  up  to  2021.09  and their  analysis  are  published  in
\citep{TESS,TESS2}. Observations  made after  that date are  posted at
the EXOFOP-TESS  web site\footnote{\url{exofop.ipac.caltech.edu}}, but
data  on the  resolved pairs  are  incomplete there  because the  main
interest   was   to  demonstrate   the   absence   of  companions   to
exohosts. However, new pairs resolved in this campaign are interesting
in  their own  right, for  example when  they host  eclipsing binaries
(i.e. are hierarchical systems) or show appreciable orbital motion.  We
include in Table~1 391 measures  of resolved TOIs made between 2021.09
and 2024.0.

\begin{figure}[ht]
\epsscale{1.1}
\plotone{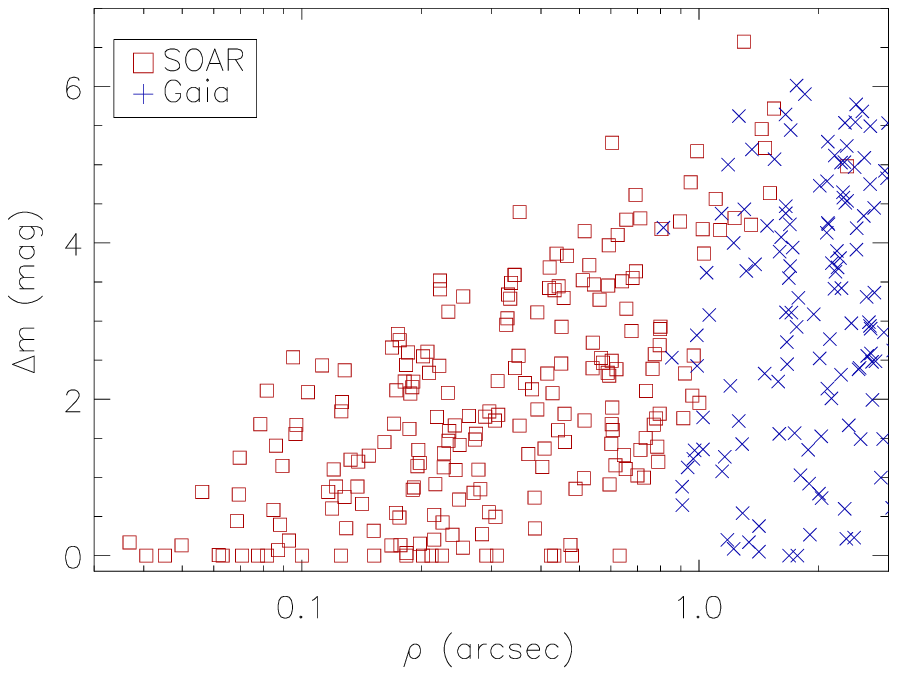}
\caption{Magnitude difference $\Delta m$ vs. separation $\rho$ for 220
  TOIs resolved  at SOAR in  2021--2024 (squares) and for 149  wider pairs
  also found in GDR3 (crosses).
\label{fig:gaia} 
}
\end{figure}

Many pairs wider than $\sim$0\farcs8  are also detected by Gaia.  They
are marked by the tag 'G'  when both components have common parallaxes
and/or PMs in  GDR3 and by the tag 'g'  otherwise (likely unrelated or
optical  pairs);  there  are 82  and  67  pairs  with  G and  g  tags,
respectively,   among   the   TOIs   resolved   before   2024.    This
G/g classification  based on  a quick  look  at the  Gaia astrometry  is
tentative.  The  remaining 220 TESS  pairs discovered at SOAR  are not
resolved by Gaia.  Figure~\ref{fig:gaia}  illustrates the detection of
binaries  by HRCam  (squares) and  by GDR3  (crosses); the  latter has
a comparable  dynamic range  for pairs  wider than  $\sim$0\farcs8.  The
Gaia resolution of  binaries is expected to improve  in the subsequent
data releases,  eventually down to  the diffraction limit  of 0\farcs1
which is still $\sim$4 times larger than at SOAR.

\begin{figure}[ht]
\epsscale{1.1}
\plotone{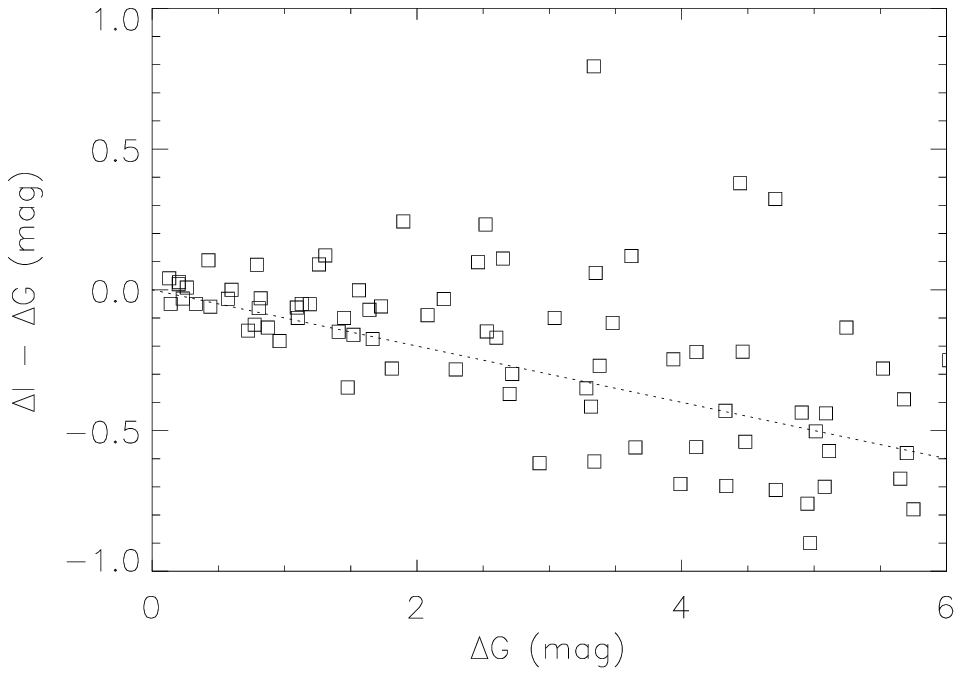}
\caption{Comparison of  the magnitude  difference $\Delta  I$ measured
  by HRCam  with $\Delta G$ measured  by Gaia for the  physical TOI
  pairs.   The dashed  line corresponds  to $  \Delta I  = 0.9  \Delta
  G$.  The outlier  (J21564+2041,  HD~208258, 2\farcs09,  $\Delta G  =
  3.34$, $\Delta I = 4.13$) may have a blue companion.
\label{fig:ptm} 
}
\end{figure}

The  TESS-Gaia  pairs  have  been discovered  at  SOAR  independently,
offering a  blind test of the  HRCam data.  The comparison  is limited
here to the  82 physical pairs where Gaia measured  both the positions
and the  magnitude differences $\Delta  G$.  The Gaia  positions match
HRCam to within 0.5\% of the  separation in both radial and tangential
directions.   The small  magnitude differences  $\Delta I$  agree with
$\Delta  G$   very  well  (Figure~\ref{fig:ptm}),  while   the  larger
differences   are  systematically   smaller   because  the   effective
wavelength in  the $I$ band is  longer than in $G$,  and the secondary
companions are redder than the primaries.   For pairs with $\Delta G <
2$ mag, the  rms scatter of $\Delta  I - \Delta G$ is  0.11 mag, while
for the  remaining pairs with  larger contrast  it is 0.35  mag. These
estimates of the  photometric accuracy of HRCam are  useful because it
is  the only  source  of  relative photometry  for  many close  pairs,
including those with  known orbits.  Masses of stars  derived from the
orbits,  parallaxes, and  relative photometry  serve to  check stellar
evolutionary models.

\subsection{Other New Pairs}
\label{sec:new}

Newly  resolved pairs  are  marked by  the  tag N  in  column (15)  of
Table~\ref{tab:measures}. There are 394  such entries referring to 353
pairs  (some pairs  were observed  more than  once), without  counting
additional 199 wider  Gaia pairs discovered at  SOAR independently and
67 new pairs  with tags A and  C.  New inner pairs  in 37 hierarchical
systems are discussed above  in Section~\ref{sec:mult}, new metal-poor
binaries  are covered  in  Section~\ref{sec:metal-poor},  and the TOIs  in
Section~\ref{sec:TESS}.  Other discoveries are reviewed here.

Continuing  the work  started  in 2023,  we observed  doubly-eclipsing
stars  discovered  by  TESS   \citep{Kostov2022},  and  resolved  seven
additional pairs.  These are massive stars at distances on the order of
1 kpc, so only a slow orbital motion is expected.  The results of this
program, published by \citet{Majewski2025}, are duplicated here in the
data tables for completeness.

New  pairs are  being discovered  consistently among  reference stars,
deemed to be  single based on the WDS and  Hipparcos and observed
  for calibrating speckle power spectra  of known tight pairs.  Here,
twelve such binaries are revealed.   The tight pair J05422$-$3032 (HIP
26862) has estimated  period of $\sim$5 yr, and its  orbital motion is
observed within 1.5 yr.

The  large  survey  of  acceleration  stars  conducted  in  2021--2022
resolved 54  pairs (tag  A) in  addition to  several wider  pairs also
resolved  in  GDR3  (tag  G).   This  sample  is  based  on  the  GDR3
astrometry,  thus avoiding  ``obvious''  binaries without  5-parameter
solutions in GDR3.   The newly resolved pairs are  either close (below
0\farcs1)  or relatively  wide ($\sim$1\arcsec)  with dim  companions.
The second group can be  contaminated by unrelated (optical) pairs, so
we reobserve them after several years  to check if the relative motion
is consistent with  physical binaries. On the other  hand, close pairs
move fast, and  for three of them  orbits based on the  SOAR data have
been  computed (J03417$-$5126,  period  3 yr;  J10048$-$3105, 6.9  yr;
J14376$-$1632, 1.7  yr).  Acceleration  pairs are  entered in  the WDS
with DDs like FR~NN. 

New  subsystems   in  wide   binaries  selected   from  the   list  by
\citet{Andrews2017}  have   tag  C;  14   of  those  are   present  in
Table~\ref{tab:measures}.  Comments on selected  new pairs follow.


\begin{figure*}[ht]
\epsscale{1.1}
\plotone{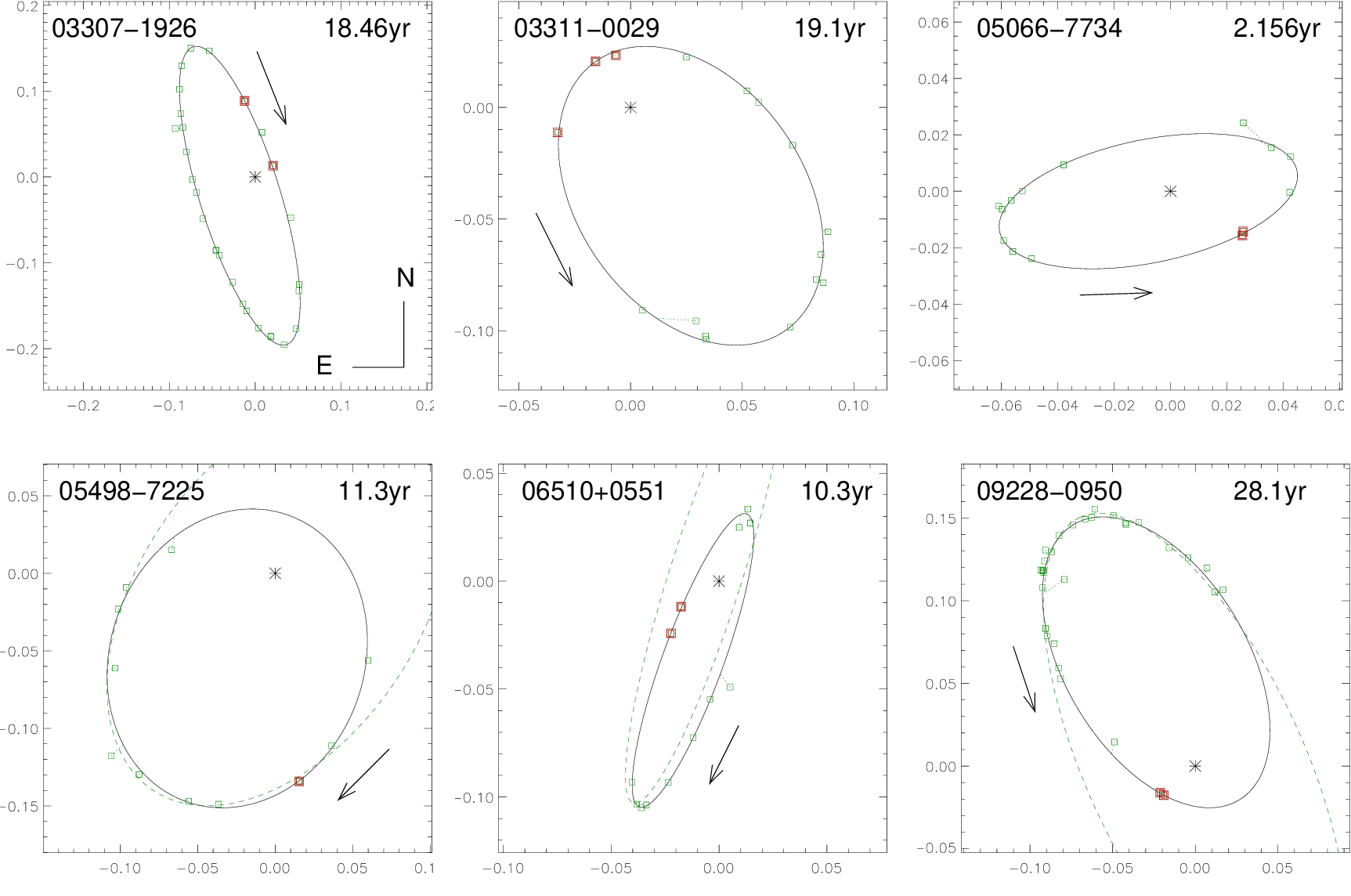}
\caption{Six  new visual  orbits with  well-constrained elements.   In
  each plot, the ellipse shows  the fitted orbit, squares connected to
  the  ellipse  are accurate  speckle  data  (bigger red  squares  for
  2024--2025), while  visual measures  and tentative speckle  data are
  plotted  as  crosses.   The   primary  component  (asterisk)  is  at
  coordinate origin, the axis scale  is in arcseconds, the orientation
  is standard (north up, east  left).  Periods and directions of motion
  are indicated.  Dashed ellipses in  the bottom row are substantially
  revised prior orbits.
\label{fig:accurate} }
\end{figure*}

{\it J00217$-$3141} (TYC 6990-325-1) is listed in the WDS as a 0\farcs8
pair TDS1418 resolved by the Tycho  satellite. This pair appears to be
spurious because  it is not seen  by either Gaia or  SOAR.  Instead, a
closer  0\farcs084  pair Aa,Ab  is  found.   Its estimated  period  is
$\sim$20 yr.

{\it J04277$-$3225}  (TYC 7038-91-1) is  a CPM companion  to HIP~20819
located at 56\arcsec.  The CPM companion is resolved as a 0\farcs7
  binary, while HIP~20819, observed as  a reference star, is found to
be single.

{\it J04365+2320} is a reference star HIP~21459, found to be double at
0\farcs17 separation with a large contrast $\Delta I = 3.6$ mag.
The estimated period of this pair is $\sim$15 yr.

{\it J04558+1502}  is another reference  star HIP~22913 found to  be a
0\farcs18 double.  However, the estimated period is about a century.

{\it  J05300+0215} (HD  287952)  is a  spectroscopic quadruple  system
detected  in  the  APOGEE  spectra  by  \citet{Kounkel2021}.   It  was
observed on suggestion by T.~Merle and found to be a 0\farcs30 pair. %

{\it J07030$-$2241} is a 2\farcs7  Tycho pair TDS4418 according to the
WDS. However, we see only a 1\farcs15  pair  with $\Delta I = 3.1$
  mag,  apparently  different  from TDS4418.   Neither  companion  is
present in  GDR3,  so we  do not know  whether the new pair  is a
  chance alignment or not.

{\it J07161$-$1431} is the reference  star HIP~35164, revealed here as
a  spectacular  0\farcs06  pair  of  equal  stars.  Its  period,  
  estimated from  the GDR3  parallax of  4.4\,mas and  separation, is
$\sim$40 yr.

{\it J07547$-$0421} was observed as TESS object of interest and found to
be a triple where the outer 1\farcs8 pair AB is present in GDR3 and
the 0\farcs26 subsystem BC is new. The GDR3 parallax of 1\,mas means
that only a very slow motion is expected. 

{\it  J08418$-$5304}  (HD 74438)  is  a  young compact  spectroscopic
quadruple in the cluster IC~2391 discovered by \citet{Merle2022}.  The
outer period, estimated to be around 5.7 yr from spectroscopy, matches
the 0\farcs034 separation, given the GDR3 parallax of 6.9\,mas.

{\it J08599$-$1546} is  a new 0\farcs04 pair, while the 0\farcs4 Tycho pair
TDS6240  is apparently spurious.

{\it J09266$-$1626} is the subdwarf  star HIP~81661 resolved in 2025 at
0\farcs1 and  showing rapid  orbital motion (estimated  period $\sim$7
yr).

{\it J12093$-$2759} (HIP 59259) is an eclipsing binary QY Hya with a
suspected tertiary companion, resolved here at 0\farcs06. The
estimated outer period is $\sim$4 yr. 

{\it  J12183$-$6301} is resolved at 0\farcs2, while the Tycho pair
TDS8311 has a 0\farcs4 separation. Pending further data, we assume the
Tycho pair to be spurious.

{\it J13044$-$1316}  is a resolved quadruple. In addition to the
known pairs AB (0\farcs5) and AC (1\farcs5), we resolve star B as a
0\farcs06 equal pair. The Ba,Bb  pair is also seen in  the 2018 SOAR data,
but it was overlooked. Its estimated period is $\sim$12 yr.

{\it  J13130$-$2430} is  clearly resolved  as a  0\farcs04 equal  pair,
while  the 0\farcs5  pair  TDS8731  appears spurious,  as 
many (but not all) similar Tycho doubles.

{\it J14025+0941}  (LP 499-23) is  revealed as a low-mass  triple Bab-C
with separations of 0\farcs08 and  0\farcs6. The outer pair RAO~309BC
has been  discovered at the 1.5  m telescope with a resolution 
insufficient for detecting the inner  subsystem; its estimated period is
7 yr. The WDS  component A at 930\arcsec ~(HD 122563)  has a similar PM
but discordant  parallax. For  this reason,  the triple  system should
have the  WDS code  14035+0935 based  on the position  of BC,  while the
optical pair AB should be ignored.

{\it  J14562+1745}  has  been  known  as a  triple  system  AB-C  with
separations  of  0\farcs1  and  1\farcs8;  we resolved  star  C  as  a
0\farcs07 pair, so this  is a 2+2 quadruple. The 43 yr  orbit of AB is
known, and  the orbit  of Ca,Cb  can be  determined quickly  (the
  estimated period of Ca,Cb is 10 yr). 

{\it J15031-4237} is a similar case where the known visual triple Aab-B
with separations of 388\arcsec ~and 0\farcs1 is turned into a quintuple
because its secondary component, L~406-116,  contains three stars in a
B-Cab  arrangement  with separations  of  0\farcs9  and 0\farcs1;  the
period of  Ca,Cb should be around  10 yr. Star B  was targeted because
GDR3 indicated multi-peak transits. All  stars in this system, located
at 40\,pc from the Sun, are dwarfs of spectral types K and M.

{\it  J15053$-$4104}  is  a  5th  magnitude  reference  star  HIP~73826
resolved tentatively at 18\,mas in the $y$ filter by elongation of its
power spectrum.  Objects observed before and after it do not show such
an  elongation.  However,  a RUWE  of 1.1  in GDR3  and a  constant RV
suggest single star, unless this pair  is a twin with $\Delta m \sim
0$. The  separation implies a  period of  $\sim$2 yr, so  double lines
might be detectable.

{\it J15436$-$8348}  (CPD-83~587), resolved at 0\farcs06,  contains a
1.47 day eclipsing binary detected by both Gaia and TESS.

{\it J15489$-$5424} (HD  140944) is a known 0\farcs5  pair FIN~61 where
we resolve  the secondary component  at 0\farcs05. Despite  its small
separation, the secondary should have a period on the order of 50 yr
because the system is located at $\sim$1 kpc distance.

{\it J18045$-$0111} is an M2 dwarf resolved at 0\farcs16. It
also has a CPM companion in GDR3 at 4\farcs6 separation. The estimated
period of Aa,Ab is $\sim$12 yr.

{\it J18280+0905.} The secondary component  in the 4\farcs4 nearby pair,
suspected to be a close binary by  the large RUWE in GDR3, is resolved
here at  0\farcs11.  GDR3  contains four stars  with similar
parallaxes  and PMs  located within  27\arcsec, so  this system  is at
least quintuple; its components have masses from 0.2 to 0.5 \msun.

{\it J18310$-$5533} is listed  in the WDS as two pairs  AB and AC with
separations of 1\farcs3 and  3\farcs8, respectively. However, the pair
AB is spurious (not seen by GDR3 and SOAR) and only AC is real. Star C
with a large RUWE in GDR3 is resolved here at 0\farcs12; its estimated
period is $\sim$20 yr.

{\it J19100+1016} is a new 0\farcs3 pair in the 3\arcsec ~Tycho double
TDT1266. The  outer companion B is  seen by GDR3, its  parallax and PM
are discordant with A, so this system is not triple.

{\it   J19127$-$3351}   is  a   triple   system   HIP~94391  of   Bab-C
architecture.  The  WDS  component   A  at  32\arcsec  ~(HIP~94393)  is
unrelated to BC.  Star A has been pointed as a reference  and resolved
into a 0\farcs22 pair.

{\it J20311$-$7508}  is yet another case where the 0\farcs4 Tycho  pair
TDS2334 turns out to be spurious, while we resolve another 0\farcs12
binary instead.  

{\it J23526+1057} (HIP 117730, 85~Peg) is a naked-eye star resolved at
0\farcs7 by \citet{Horch2008} and measured by  this team several
times using  4 m  telescopes.  The 0\farcs06
subsystem  Ba,Bb is resolved  here for  the  first time.   Its period  is
estimated to be around 4 yr. The motion of the outer pair AB which has
closed down  to 0\farcs23 can be  described by a tentative  orbit with
a 157 yr period.

\subsection{New and Updated Orbits}
\label{sec:orbits}

\begin{figure}[ht]
\epsscale{1.0}
\plotone{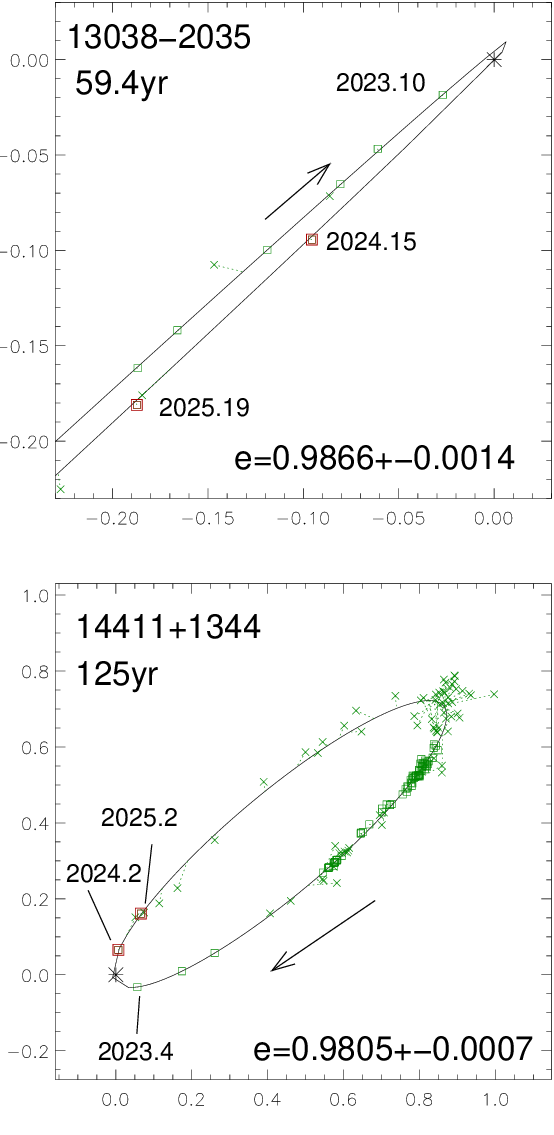}
\caption{Two   new  orbits   with   large   and  accurately   measured
  eccentricity. The eccentricity and its error are indicated.
\label{fig:eccentric} }
\end{figure}

\begin{figure*}[ht]
\epsscale{1.1}
\plotone{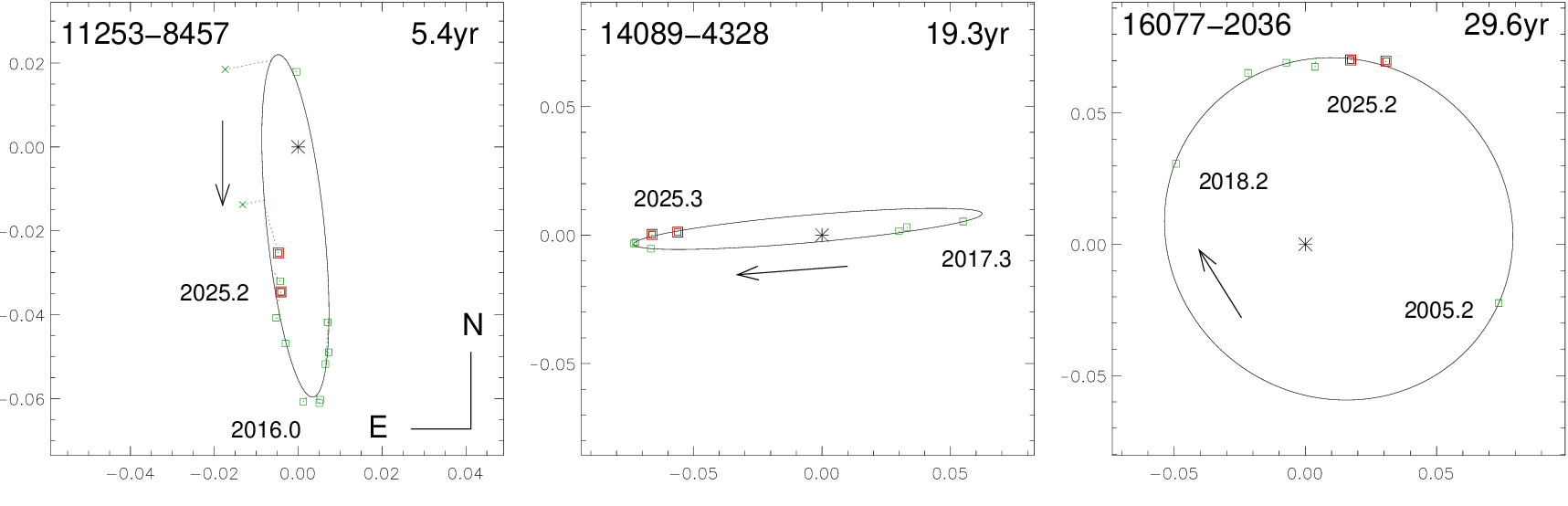}
\caption{Orbits of three young binaries (see text).
\label{fig:PMS} }
\end{figure*}

The  stream of  new measures  from HRcam  allows improvement  of known
orbits and calculation of new ones. Speckle data gradually replace the
 less  precise  and  less  reliable  visual  measures,  so  even
well-established  visual  orbits  can  be  substantially  improved  if
covered by  speckle interferometry.  The methods  of orbit calculation
and various caveats are covered  in \citet{Orbits2024} and we refer to
this publication for details such as weights and software tools. 
  Briefly,  we fit  the  orbital elements  to the  data  by the  least
  squares  method.   Some elements  are  fixed  when they  are  poorly
  constrained or degenerate. 

Orbit improvement in response to  new measures is an almost continuous
process.   Most  corrections  are   incremental  and  do  not  justify
publication   of  updated   elements,   although   the  threshold   of
``significant'' corrections is subjective.  In some cases, orbits that
appeared to be well-defined required  dramatic revisions to fit recent
measures, and  such revisions  definitely deserve publication.  On the
other hand, poorly constrained orbits  based on scarce data, useful in
their own right, are not reliable  and may be substantially revised in
the future.

New and corrected  orbits resulting from the HRCam data  appear in the
Information Circulars of the IAU Commission G1 compiled by J.A.~Docobo
at       Observatorio       Astron\'omico      Ram\'on       Mar\'{i}a
Aller.\footnote{https://www.usc.gal/astro/circularing.html}      These
orbits  are added  to  the  online ORB6  orbit  catalog maintained  at
USNO\footnote{https://crf.usno.navy.mil/wds-orb6/} \citep{VB6} and are
also featured in this series. Reliable orbits from the Circulars 214--217
are published here, with some adjustments  where necessary.

For this  paper, we selected   202 orbits where all  elements are
constrained  by the  data and  their  errors determined  by the  least
squares fit  are meaningful; they  are grades A or  B in the  sense of
\citet{Orbits2024}.    The  updates   of  known   orbits  are   deemed
substantial    or    dramatic     to    warrant    publication.     In
Table~\ref{tab:vborb}, each system  is identified by its  WDS code and
the DD  (when available), followed  by the seven Campbell  elements in
standard notation.  The two  rightmost columns contain the provisional
orbit grade  assigned according to the  ORB6 rules and a  reference to
the most current  orbit which has been here  improved, when available;
the first-time orbits are referenced as SOAR2025.
Figure~\ref{fig:accurate}  gives  illustrative  plots of six
orbits with accurate elements.

Table~\ref{tab:vborb} contains 10 orbits with eccentricities exceeding
0.9. In two such orbits the eccentricity is fixed  because the periastron
has  not been  covered,  but the  remaining  large eccentricities  are
measured accurately  owing to the  dense periastron coverage  at SOAR;
Figure~\ref{fig:eccentric} shows two  examples. Both are classical
visual pairs with relatively long periods that recently passed through
the periastron.  The  first one, J13038$-$2035  (BU 341), was  resolved at
33\,mas separation  in 2023.10, unresolved in  2023.42, passed through
the periastron  in  2023.67 (when  it  was  not  visible from  SOAR),  and
reappeared again  in 2024.16  at 0\farcs13,  opening up  on its  59 yr
orbit. The eccentricity of $0.9866 \pm 0.0014$ is probably the largest
reliably  measured one,  differing  from unity  by  only 0.0134.   The
second  pair  J14411+1344  (STF1965AB) passed  through the periastron  in
1898.4  and in  2023.94;  the  last periastron  was  covered at  SOAR,
constraining  the eccentricity  to $e  =  0.9805 \pm  0.0007$ in  this
otherwise well-characterized orbit of grade 1 with a period of 125 yr.

In J13038$-$2035 (HD 113415),  the separation between two equal
solar-type stars  at periastron is  only 0.27 au.  Twin  binaries with
orbits  of  $\sim$20  au  are  believed to  form  via  accretion  from
circumbinary disks accompanied  by migration \citep{Offner2023}, which
should have damped the  eccentricity.  Hence, the extreme eccentricity
must have been acquired later, when the stars were already on the main
sequence  and  the  gas  infall stopped.   A  plausible  mechanism  is
disruption of  dynamically unstable triples.  Both  binaries mentioned
here lack known outer companions, supporting the disruption scenario.
  
Orbits of young  binaries are valuable for measuring  their masses and
testing evolutionary  models \citep{Simon2019}.   With this goal
in mind,  many young pairs  are being  monitored at SOAR.   Three such
orbits   are   shown   in  Figure~\ref{fig:PMS}.    The   tight   pair
J11253$-$8457 (HIP~55746, BRC  4Aa,Ab) has been discovered  at SOAR in
2016    in    a    survey   of    the    $\epsilon$~Cha    association
\citep{Briceno2017}.  Its preliminary 9 year circular orbit is updated
here to  5.4 year period. The  data cover almost two  full cycles, but
the separation never exceeds 61\,mas,  so larger telescopes are needed
to map  the orbit  near periastron. Remarkably,  $\epsilon$~Cha itself
was revealed  at SOAR as  a close triple, and  the orbit of  its inner
subsystem is  now known \citep{TRI23}.  The  second pair J14089$-$4328
(HIP  69113,  HJ~4653Aa,Ab)  of  spectral  type  B9V  belongs  to  the
Scorpius-Centaurus (Sco-Cen)  association; it  has been  identified at
SOAR in 2017.3,  and its first 19.3 yr orbit  is determined here.  The
first orbit  of J16077$-$2036 (BOY~20),  also a member of  Sco-Cen, is
based  on the  SOAR data,  apart from  the 2005  discovery measure  by
\citet{Bouy2006}.  The inclination is fixed  at 160\degr ~to avoid the
degenerate face-on solution.  Half of the 29 yr orbit is covered.  Two
similar   M4Ve   stars  comprising   this   pair   are  variable   and
chromospherically active.  The mass sum  of 1.14 \msun (using the GDR3
parallax of 6.86$\pm$0.07 mas) indicates  that these stars will become
late-K dwarfs when they settle on the main sequence.





\begin{deluxetable*}{l l cccc ccc cc}    
\tabletypesize{\scriptsize}     
\tablecaption{Visual Orbits
\label{tab:vborb}          }
\tablewidth{0pt}                                   
\tablehead{                                                                     
\colhead{WDS} & 
\colhead{Discoverer} & 
\colhead{$P$} & 
\colhead{$T$} & 
\colhead{$e$} & 
\colhead{$a$} & 
\colhead{$\Omega$ } & 
\colhead{$\omega$ } & 
\colhead{$i$ } & 
\colhead{Grade }  &
\colhead{Reference\tablenotemark{a}} \\
\colhead{$\alpha$, $\delta$ (2000)} &
\colhead{Designation} &  
\colhead{(yr)} &
\colhead{(yr)} & &
\colhead{(arcsec)} & 
\colhead{(deg)} & 
\colhead{(deg)} & 
\colhead{(deg)} &  & 
}
\startdata
00111$+$0513 & TOK 869 & 8.377 & 2017.893 & 0.1648 & 0.1079 & 111.6 & 223.9 & 48.9 & 2 & Tok2023a \\
             &     & $\pm$0.077 & $\pm$0.132 & $\pm$0.0109 & $\pm$0.0019 & $\pm$1.6 & $\pm$5.3 & $\pm$1.5&     &  \\
00261$-$1123 & YR 4 & 42.13 & 2025.963 & 0.6301 & 0.2538 & 49.8 & 30.2 & 166.4 & 3 & Msn2023 \\
             &     & $\pm$1.25 & $\pm$0.055 & $\pm$0.0128 & $\pm$0.0037 & $\pm$19.3 & $\pm$20.8 & $\pm$7.8&     &  \\
00417$-$2446 & B 10 & 171. & 2023.428 & 0.990 & 0.2523 & 44.8 & 69.1 & 109.0 & 3 & SOAR2025 \\
             &     & $\pm$32. & $\pm$0.171 & fixed & $\pm$0.0306 & $\pm$3.8 & $\pm$2.3 & $\pm$3.5&     &  \\
00460$-$3043 & HDS 100 AB & 35.2 & 2024.054 & 0.7560 & 0.1238 & 33.0 & 255.2 & 104.7 & 3 & Tok2023e \\
             &     & $\pm$5.0 & $\pm$0.054 & $\pm$0.0245 & $\pm$0.0129 & $\pm$1.2 & $\pm$3.1 & $\pm$0.9&     &  \\
01055$+$1523 & GKM9017 Ba,Bb & 2.8767 & 2024.476 & 0.270 & 0.0812 & 54.7 & 270.2 & 95.9 & 3 & SOAR2025 \\
             &     & $\pm$0.0256 & $\pm$0.067 & $\pm$0.082 & $\pm$0.0056 & $\pm$1.0 & $\pm$4.8 & $\pm$1.2&     &  \\
\enddata 
\tablenotetext{a}{Orbit References are provided at
  \url{https://crf.usno.navy.mil/data_products/WDS/orb6/wdsref.html} }
(This table is available in its entirety in machine-readable form in the online article.)
\end{deluxetable*}

Eight orbits use both positional measures and radial velocities (RVs).
In  addition  to  the  visual elements  in  Table~\ref{tab:vborb},  we
provide in Table~\ref{tab:sborb} the RV  amplitudes of the primary and
secondary components $K_1$ and $K_2$  and the systemic velocity $V_0$.
These parameters, delivered by the joint fit of positions and RVs, are
similar but  not equal to  the published spectroscopic  elements.  The
first three  columns repeat the WDS  codes, DDs, and periods,  and the
last column gives  references to publications containing  the RVs used
here.   Unpublished RVs  measured  at  the 1.5  m  CTIO telescope  are
referenced  as  CHIRON \citep{CHI25}.   In  the  combined orbits,  the
angles $\Omega$ and $\omega$ are  selected to match both the ascending
node and RV of the primary component and the relative positions of the
secondary component.

\begin{deluxetable*}{l l cc cc l}    
\tabletypesize{\scriptsize}     
\tablecaption{Combined Spectro-Interferometric Orbits
\label{tab:sborb}          }
\tablewidth{0pt}                                   
\tablehead{                                                                     
\colhead{WDS} & 
\colhead{Discoverer} & 
\colhead{$P$} & 
\colhead{$K_1$} & 
\colhead{$K_2$} & 
\colhead{$V_0$} &
\colhead{Reference}  \\
\colhead{$\alpha$, $\delta$ (2000)} &
\colhead{Designation} & 
\colhead{(yr)} &
\colhead{(\kms)} &
\colhead{(\kms)} &
\colhead{(\kms)} & 
}
\startdata
04258$+$1800 & COU2682 & 45.880 & 2.43 & \ldots & 39.96 &  \citet{Griffin2012} \\ 
             &   & $\pm$0.358 & $\pm$0.07 & \ldots & $\pm$0.05 &  \\
05066$-$7734 & TOK 875 & 2.156 & 12.16 & 12.97 & 27.46 &  CHIRON \\
             &   & $\pm$0.001 & $\pm$0.02 & $\pm$0.02 & $\pm$0.01 &  \\
11418$+$0508 & TOK 896 & 3.650 & 5.73 & 7.71 & 18.25 &  \citet{Sperauskas2019} \\ 
             &   & $\pm$0.005 & $\pm$0.16 & $\pm$0.27 & $\pm$0.14 &  \\
13344$-$2730 & TOK 898 Aa,Ab & 9.5874 & 2.03 & \ldots & $-$73.28 &  \citet{Barbato2023} \\ 
           &   & $\pm$0.0203 & $\pm$0.01 & \ldots & $\pm$0.01 &  \\
17433$+$2137 & DUQ 1 Aa,Ab    & 6.9621  & 9.10 & \ldots & 23.85 &  \citet{Duquennoy1996} \\ 
             &   & $\pm$0.0202 & $\pm$0.12 & \ldots & $\pm$0.04 &  \\
19126$+$1651 & WSI 107 Ca,Cb & 6.993 & 4.61 & \ldots & 34.09 & \citet{Griffin1979} \\ 
             &   & $\pm$0.019 & $\pm$0.25 & \ldots & $\pm$0.12 &  \\
19155$-$2515 & B 430 & 19.968 & 12.21 & 11.55 & $-$26.80 &  \citet{Fekel1975} \\ 
             &   & $\pm$0.031 & $\pm$1.28 & $\pm$1.48 & fixed &  \\
19512$-$7248 & TOK 697 Aa,Ab & 2.9655 & 11.67 & 13.82 & 7.21  & CHIRON \\
             &   & $\pm$0.0178 & $\pm$0.96 & $\pm$0.85 & $\pm$0.24 &  \\
\enddata 
\end{deluxetable*}

\subsection{Spurious Pairs}
\label{sec:bogus}


\begin{deluxetable*}{ l l l l  } 
\tabletypesize{\scriptsize}    
\tablecaption{Likely Spurious Pairs
\label{tab:bogus} }                   
\tablewidth{0pt}     
\tablehead{ \colhead{WDS}  &
\colhead{Discoverer}  &  
\colhead{Resolved} & 
\colhead{Unresolved\tablenotemark{a}} 
}
\startdata  
00397$-$2628 & HIP 3125 & 0.25 Spe 2021.7 & 2023--24 R \\ 
02054$-$0947 & LSC 131Aa,Ab & 0.2 Spe 2019 & 2024.9 R \\ 
04469$-$6036 & TOK 388AB & 0.04 Spe 2014 & 2014-2024 R,S  \\ 
09098+1134 & CHR 131 & 0.09 Spe 1986 & 2024 R,S \\ 
\enddata
\tablenotetext{a}{Additional indications of the spurious nature of visual pairs:
R -- no excess noise in GDR3, RUWE$<$2; 
L -- long estimated period; 
S -- short estimated period or spectroscopic coverage;
dm -- discrepant magnitude difference.
}
 (This table is available in its entirety in machine-readable form in the online article.)
\end{deluxetable*}

Some  double stars  listed  in  the WDS  are  actually single.   These
spurious  pairs originate  from erroneous  visual resolutions  or
  dubious speckle  data caused by instrumental  artifacts like optical
  ghosts shown in Figure~11 of TMH10 and in Figure~3 of \citep{SAM17}.
 Real binary stars, e.g. close  ones, also can be resolved spuriously
for the same reasons.  Identifying  spurious pairs will save observing
time in  the future by  eliminating the  need to followup  and examine
these  targets.   In  Table~\ref{tab:bogus}  are listed  49  pairs  we
identify as likely  spurious, continuing the clean-up  effort from our
previous papers. In  this table we provide the WDS  identifier and DD,
the separation,  method, and date  of the original discovery,  and the
year(s) it has  been unresolved in this program.  Following  that is a
code giving  other indications supporting the  characterization of the
double  as  spurious  \citep[see  details in][]{SAM21}.   In  the  WDS
\citep{WDS},  these pairs  are not  removed but  are given  an X  code
identifying them as a ``dubious double'' or a ``bogus binary''.

Among the WDS binaries within  100\,pc, 38 pairs were unresolved.  All
these stars  have parallaxes  in GDR3. The WDS  pairs with  multiple prior
observations are likely real, but either are too close or have a large
contrast.   On  the  contrary,  most occultation  binaries  have  been
detected  only  once and  never  confirmed,  so they  are  suspicious.
Another  known  source  of  spurious  pairs  with  separations  around
0\farcs5 is  the Tycho  double star  catalog (DD  codes TDS  and TDT),
although several such pairs are confirmed  here and one is detected as
triple (Section~\ref{sec:mult}). WDS J21549$-$7720  is a real 7\farcs6
pair  detected by  both  Gaia  and SOAR;  its  separation of  2\farcs1
measured  by  Hipparcos  (HDS3117AB)  is  likely  explained  by  wrong
interpretation of the modulation  produced its focal-plane grating, as
happened in a few other cases. The HDS3117AB pair is thus spurious.

\section{Summary}
\label{sec:sum}

This paper reports  results of the multi-year  campaign of double-star
observations  with  the  SOAR   speckle  camera,  continuing  previous
publications  in this  series.   A total  of  7299 observations  (both
measures and  non-resolutions) obtained in  2024 and 2025, as  well as
previously  unpublished earlier  data, is  the core  of this  work. We
resolved for  the first time  over 400 pairs,  including new
inner subsystems  in triple stars within  100\,pc, metal-poor binaries
for future mass measurement, and TOIs. The observations were organized
by combining  regular time allocations  from several projects  and the
engineering time into a single  program. The benefits of this approach
are the efficient use of telescope time, the improved time coverage of
pairs with  fast motion, and the  unified data reduction  and calibration.

The    SOAR   speckle    data   are    used   in    several   research
projects. Calculation of orbits is the obvious application, leading in
turn  to  the  characterization  of orbital  dynamics  in  binary  and
higher-order system and to the measurement of stellar masses. The work
by \citet{Vrijmoet2026}  on the  orbits of  54 nearby  M dwarfs  is an
excellent illustration  of this  point. The  median semimajor  axis of
those binaries is  0\farcs14, and the median orbital period  is 6.4 yr.
Without a  dense coverage  (impossible under the classical time  allocation
scheme), most orbits would have remained undetermined.

The   SOAR    speckle   program    complements   the    Gaia   mission
(2014--2025). Gaia data  are used here for  selecting likely binaries,
for estimation  of periods  from separations  and parallaxes,  and for
measurement of masses.  The angular resolution of SOAR exceeds that of
Gaia by  a factor  of four, so  even in the  final Gaia  data release,
still years  ahead, many  close binaries  will remain  unresolved. The
SOAR  data  will help  in  determination  of orbital  and  astrometric
parameters  via joint  solutions  using both  Gaia  transits and  SOAR
relative  positions.   For  triple   and  higher-order  systems,  Gaia
pipelines often fail  owing to complex nature of  the signals.  Future
demand  for high  angular resolution  data will  be largely  driven by
Gaia, and SOAR is posed to play a pivotal role here.

\begin{acknowledgments} 

We  thank SOAR  director  C.~Brice\~no for  allocating some  technical
time.  E.C.  and  R.A.M acknowledge support form  FONDECYT/ANID \# 124
0049. R.A.M  also acknowledges support from  Fondo GEMINI, Astr\'onomo
de Soporte GEMINI-ANID grant \# 3223  AS0002. The research of A.T.\ is
supported by the NSFs NOIRLab.   Comments by the anonymous referee
  helped to improve the presentation. 

This work used the SIMBAD service operated by Centre des Donn\'ees Stellaires 
(Strasbourg, France), bibliographic references from the Astrophysics Data System
maintained by SAO/NASA, and the Washington Double Star Catalog maintained at the
USNO. This work has made use of data from the European Space Agency (ESA) 
mission Gaia (\url{https://www.cosmos.esa.int/gaia}) processed by the Gaia Data
Processing and Analysis Consortium (DPAC, {\url 
https://www.cosmos.esa.int/web/gaia/dpac/consortium}). Funding for the DPAC has 
been provided by national institutions, in particular the institutions 
participating in the Gaia Multilateral Agreement.
\end{acknowledgments}

\facility{SOAR}


\bibliography{soar.bib}
\bibliographystyle{aasjournal}

\appendix

\section{SOAR calibrators}
\label{sec:calmodels}

In  this Appendix,  we give  the updated  set of  the SOAR  calibrator
binaries and the models of their motion. Table~\ref{tab:cal} lists the
WDS codes, DDs,  and Hipparcos numbers of 86  binaries.  The following
columns contain the  number of runs $N$ (multiple  observations in one
run, if available, are averaged and considered as single measure), the
mean  separation  $\rho$,  and  the  approximate  magnitude  difference
$\Delta m$.   The linear and  orbital models are  denoted by 1  and 2,
respectively. The  last two  columns give the  rms residuals  from the
models in  tangential and radial  directions. As noted, a  few deviant
(by  more   than  2.5$\sigma$)  measures  of   some  calibrators  were
discarded.


\begin{deluxetable*}{l l c c   c c  c  cc } 
\tabletypesize{\scriptsize}
\tablewidth{0pt}
\tablecaption{List of Calibrator Binaries 
\label{tab:cal}}
\tablehead{
\colhead{WDS} &
\colhead{DD} &
\colhead{HIP} &
\colhead{$N$ } &
\colhead{$\rho$ } &
\colhead{$\Delta m$ } &
\colhead{Mod.} &
\colhead{$\sigma_{\rm tan}$ } &
\colhead{$\sigma_{\rm rad}$ } \\
 &   &  &  &
\colhead{(\arcsec)} &
\colhead{(mag)}  & &
\colhead{(mas)} &
\colhead{(mas)} 
}
\startdata 
00098$-$3347& SEE 3      &     794& 12  &  0.78 & 1.47 & 2  & 1.7 &  2.1 \\
00522$-$2237& STN 3AB    &    4072& 31  &  1.99 & 0.78 & 1  & 2.0 &  3.5 \\
01024$+$0504& HDS 135AB  &    4849& 11  &  0.71 & 1.36 & 2  & 3.0 &  1.4 \\
\enddata
\tablecomments{(This table is available in its entirety in machine-readable form in the online article.)}
\end{deluxetable*}

\begin{deluxetable}{l  c c c c c } 
\tabletypesize{\scriptsize}
\tablewidth{0pt}
\tablecaption{Linear Models \label{tab:lin}}
\tablehead{
\colhead{WDS} &
\colhead{$t_0$} &
\colhead{$\theta_0$} &
\colhead{$\dot{\theta}$} &
\colhead{$\rho_0$} &
\colhead{$\dot{\rho}$} \\
& \colhead{(JY)} &
\colhead{(\degr)} &
\colhead{(\degr ~yr$^{-1}$)} &
\colhead{(\arcsec)} &
\colhead{(\arcsec ~yr$^{-1}$)} 
}
\startdata
00522$-$2237 &   2019.989 &  241.015 &   $-$0.243 &   1.9832 &   0.0005 \\
01037$-$3024 &   2019.193 &  236.870 &   $-$0.106 &   0.8563 &   0.0022 \\
05019$-$7638 &   2020.356 &  154.993 &    0.634 &   0.8251 &   0.0037 \\
05508$-$2907 &   2019.420 &  329.875 &    0.087 &   2.1657 &   0.0019 \\
06425$-$4234 &   2019.055 &  194.781 &    0.151 &   2.3363 &  $-$0.0005 \\
08461$+$0748 &   2020.828 &  339.942 &    0.033 &   2.7972 &   0.0016 \\
09125$-$4337 &   2019.517 &  280.826 &   $-$0.009 &   2.8983 &   0.0037 \\
13142$-$1634 &   2021.287 &  259.584 &   $-$0.029 &   1.6688 &   0.0112 \\
15308$-$0746 &   2021.109 &  342.180 &    0.883 &   1.0561 &   0.0024 \\
15428$-$1601 &   2020.740 &  110.771 &    0.159 &   1.5931 &  $-$0.0102 \\
17315$-$4156 &   2021.636 &   37.106 &   $-$0.160 &   1.4509 &   0.0040 \\
19426$-$5901 &   2020.536 &  148.749 &   $-$0.119 &   2.4728 &   0.0099 \\
22152$-$0535 &   2018.233 &  280.213 &    0.015 &   0.6954 &   0.0001 \\
\enddata
\end{deluxetable}

The  slow  motion of  a  wide  binary  can  be represented  by  linear
functions of time in polar coordinates ($\rho$, $\theta$):
\begin{eqnarray}
\theta(t) & = & \theta_0 + \dot{\theta} \; (t-t_0) , \\
\rho(t) & = & \rho_0 + \dot{\rho} \; (t-t_0), 
\end{eqnarray}
where $t_0$ is the mean  time of observation. These equations describe
a spiral  trajectory.  Fragment of  a spiral is  a better match  to an
orbital segment  than a linear trajectory  in rectangular coordinates,
with the  same number  of 5  free parameters  (for a  circular face-on
orbit, the spiral is accurate).  The  choice of $t_0$ ensures that the
offsets and slopes are statistically independent. Parameters of the
linear models for 13 calibrators are given in Table~\ref{tab:lin}.

Table~\ref{tab:orbcal} lists the orbital  elements of the remaining 73
pairs, in  standard notation. The  last column gives reference  to the
catalog orbits used as templates (the reference codes are available in
ORB6).  Only  the elements $(T,  a, \Omega)$  were fitted to  the SOAR
data (including also  Hipparcos and GDR3 positions  when available) if
the  templates are  adequate.  Otherwise,  all elements  were slightly
corrected   using   all   available   data   with   suitable   weights
\citep{Orbits2024}, and  then the 3  elements were fitted to  the SOAR
data only;  such cases are  marked by asterisks after  the references.
Cases  when corrections  of the  template orbits  are substantial  are
marked by  single asterisks in  the reference column. We  caution that
the orbits  in Table~\ref{tab:orbcal} are specially  adjusted to match
the  SOAR data  and  are  not necessarily  the  best  orbits of  these
binaries in general sense (although  they might be better than current
catalog   orbits).    There  are   five   overlaps   with  orbits   in
Table~\ref{tab:vborb} (WDS  codes 04199+1631,  10426+0335, 14489+0557,
16044$-$1122,  and   23100$-$4252);  Table~\ref{tab:vborb}   should  be
preferred over the orbital models of the calibrators.


\begin{deluxetable*}{l  ccc ccc c  l } 
\tabletypesize{\scriptsize}
\tablewidth{0pt}
\tablecaption{Orbital Elements 
\label{tab:orbcal}}
\tablehead{
\colhead{WDS} &
\colhead{$P$} &
\colhead{$T$} &
\colhead{$e$} &
\colhead{$a$} &
\colhead{$\Omega$} &
\colhead{$\omega$} &
\colhead{$i$} & 
\colhead{Reference\tablenotemark{a}}\\
& \colhead{(yr)} &
 \colhead{(JY)} & &
\colhead{(\arcsec)} &
\colhead{(\degr)} &
\colhead{(\degr)} &
\colhead{(\degr)} &
}
\startdata
00098$-$3347 &      295.106 &    1978.59 &   0.7668 &   0.9242 &   275.70 &    71.15 &    32.83 & Tok2024a* \\
01024$+$0504 &       28.320 &    2002.70 &   0.6721 &   0.4616 &   268.23 &    22.08 &   144.79 & Tok2015c* \\
01084$-$5515 &      289.600 &    1920.48 &   0.5970 &   1.1863 &    24.44 &   292.20 &    71.20 & Lin2019a \\
01158$-$6853 &       85.137 &    2001.26 &   0.0396 &   1.0868 &   140.67 &   132.65 &    31.08 & Tok2024a \\
01262$-$6751 &      374.193 &    1951.95 &   0.8491 &   1.1875 &   135.82 &   347.65 &    47.47 & Izm2019 \\
\enddata
\tablenotetext{a}{Orbit References are provided at
  \url{https://crf.usno.navy.mil/data_products/WDS/orb6/wdsref.html}.  The asterisks mark references where more than three elements are
  adjusted. Substantial orbit revisions have single asterisks in the Reference column. 
 }
(This table is available in its entirety in machine-readable form in the online article.)
\end{deluxetable*}



\end{document}